\documentclass[preprint,showpacs,showkeys,preprintnumbers,amsmath,amssymb,aps]{revtex4-1}

\usepackage{bm,amssymb,amsmath,mathrsfs}
\usepackage[usenames]{color}
\usepackage{hyperref}

\usepackage{dcolumn}
\usepackage[pdftex]{graphicx}

\newcommand{\bc}{\begin{center}}
\newcommand{\ec}{\end{center}}
\newcommand{\bit}{\begin{itemize}}
\newcommand{\eit}{\end{itemize}}
\newcommand{\bq}{\begin{equation}}
\newcommand{\eq}{\end{equation}}
\newcommand{\pprime}{{\prime\prime}}

\begin{document}

\title{Spin decoherence rate in a weak focusing all-electric ring}

\author{S.~R.~Mane}
\email{srmane001@gmail.com}

\affiliation{Convergent Computing Inc., P.~O.~Box 561, Shoreham, NY 11786, USA}

\begin{abstract}
I previously derived the expression for the spin decoherence rate 
(for orbital and spin motion in the horizontal plane)
in a weak focusing all-electric storage ring with a radial field (logarithmic potential, field index $n=0$).
Here I generalize the calculation to an arbitrary field index $n\ge0$.
I also solve the model for the relativistic Kepler problem (field index $n=1$),
where the solution for the orbit is known analytically, and I verify that it confirms the solution from perturbation theory.

{\em Version 2: I solved the model also for off-energy dispersion orbits.
The dispersion orbits are circles and I calculated the spin decoherence rate exactly.
}

{\em Version 3: I solved the model also for vertical motion, for $n=0$ (logarithmic potential).
I have also solved for the spin precession the Kepler problem (inverse square law) exactly; the paper is in press.
Extra appendices have been added;
in particular, Ivan Koop kindly gave permission to reproduce his elegant analysis of motion in a vertical spiral in a logarithmic potential.
}

{\em Version 4: Added synchrotron oscillations.}

\end{abstract}



\maketitle

\section{\label{sec:intro} Introduction}
I derive an expression for the spin decoherence rate, for motion in a homogeneous weak focusing all-electric storage ring.
Fundamentally, the use of a homonegeous smooth focusing model simplifies the analysis to that of 
relativistic charged particle motion in a central field of force.
This is a standard problem in textbooks of classical mechanics,
at least for the orbital motion (in undergraduate texts the motion is typically nonrelativistic).
A central force renders the problem analytically tractable.
In some cases, I derive the exact solution.
The derivation of analytical formulas, for a simplified model,
furnishes precise benchmark tests for numerical tests such as tracking simulations.
The analytical results may also be useful for numerical estimates for more complicated ring designs.

In \cite{ManeElectrostaticBending2012},
I derived the relevant orbital and spin equations of motion 
for rings with arbitrary electrostatic and magnetostatic fields (and rf cavities).
Some important work on the orbital motion in all-electric rings had already been derived much earlier by 
Laslett \cite{Laslett_elec_ring}.
I previously solved the problem of the spin decoherence rate
for horizontal orbits in a logarithmic potential (field index $n=0$) \cite{ManeArXivCC14_5};
here I solve the same problem for an arbitrary field index $n\ge0$.
I shall derive the expression for the spin decoherence rate,
for both off-energy dispersion orbits and horizontal betatron oscillations.
For the off-energy motion, the dispersion orbits are circles and I shall solve the problem exactly.
For the betatron oscillations, I shall employ perturbation theory and canonical transformations.
I have also derived the exact solution for the spin precession in the 
relativistic Kepler problem (inverse square force law, field index $n=1$),
where the solution for the orbit is known analytically; the solution is in press \cite{ManeRelKepler}.
I also briefly treat the spin decoherence due to vertical orbital motion, in some appendices.
In particular, Ivan Koop kindly gave permission \cite{Koopprivcomm}
to reproduce his elegant analysis of motion in a vertical spiral in a logarithmic potential;
see Appendix \ref{sec:koopsoln}.
I derive the exact solution for the orbit and spin motion in a vertical spiral in a logarithmic potential,
in Appendix \ref{sec:logpot}.

I also calculated the spin decoherence rate for synchrotron oscillations (without betatron oscillations).
To do so I performed a canonical transformation to separate the radial motion into betatron and dispersion terms.
This required the determination of the second order dispersion.
I then diagonalized the Hamiltonian using action-angle variables to determine 
the relevant averages for the synchrotron oscillations.
The results are in good agreement with those from tracking simulations.

Also not least, the analysis of a simple model clarifies fundamental principles 
of the orbital and spin motion and the spin decoherence rate.
For example, the analysis by Talman and Talman \cite{TalmanIPAC2012}
contains some subtle misconceptions and quantitative errors,
which are also discussed in the appendices.

\section{\label{sec:basic} Basic notation and definitions}
\subsection{General remarks}
I treat a particle of mass $m$ and charge $e$,
with velocity $\bm{v}=\bm{\beta}c$ and Lorentz factor $\gamma=1/\sqrt{1-\beta^2}$.
I shall mostly set $c=1$ below.
The particle spin $s$ is treated as a unit vector and $a=\frac12(g-2)$ denotes the magnetic moment anomaly.
There is no magnetic field in the model I treat; the ring is all-electric.
I shall treat a homogeneous weak focusing model below, so the reference orbit is a circle of radius $r_0$.
The arc-length along the reference orbit is $s$ and the azimuth is $\theta = s/r_0$.
The radial coordinate is $x$ and $r=r_0+x$ and the vertical coordinate is $z$.
The Hamiltonian with $s$ as the independent variable is
\bq
\label{eq:kham}
K = -p_s = -\frac{r}{r_0}\,\biggl[\, \frac{(H-\Phi)^2}{c^2} - m^2c^2 - p_x^2 - p_z^2 \,\biggr]^{\frac12} \,.
\eq
Here $H$ is the total energy. It a dynamical invariant in this model.
Note that $H = \gamma mc^2 + \Phi$.
For off-energy orbits,
following \cite{ManeElectrostaticBending2012},
I shall employ the parameter $\lambda_p = (1/\beta_0^2)(\Delta H/H_0)$.
The Thomas-BMT equation \cite{Thomas,BMT} for spin motion in prescribed external 
electric and magnetic fields $\bm{E}$ and $\bm{B}$, respectively, is
\bq
\label{eq:tbmt}
\frac{d\bm{s}}{dt} =
-\frac{e}{mc}\,\biggl[\biggl(a+\frac{1}{\gamma}\biggr)\,\bm{B}
-\frac{a\gamma}{\gamma+1}\,\bm{\beta}\cdot\bm{B}\,\bm{\beta}
-\biggl(a+\frac{1}{\gamma+1}\biggr)\,\bm{\beta}\times\bm{E}\,\biggr]\times\bm{s} \,.
\eq
As stated above, I shall set $\bm{B}=0$ below and treat only motion in an electrostatic field.
In the absence of a magnetic field, 
the equation of motion for the helicity $\bm{s}\cdot\hat{\bm{\beta}}$ is given by \cite{MSY2}
\bq
\label{eq:helicityefield}
\frac{d\ }{dt}(\bm{s}\cdot\hat{\bm{\beta}}) = \frac{e}{mc}\biggl(a - \frac{1}{\beta^2\gamma^2}\biggr)
(\bm{\beta}\times\bm{E})\cdot(\bm{s}\times\hat{\bm{\beta}})\,.
\eq
Note that the right-hand side vanishes when $a = 1/(\beta^2\gamma^2)$; this is referred to as the `magic' condition.
Hence the literature employs the terms `magic gamma' and `magic momentum' which I shall use below. 
In the model I treat, the reference orbit is at the magic gamma, hence  $a = 1/(\beta_0^2\gamma_0^2)$.

\subsection{Motion in horizontal plane}
For most of this paper, I shall restrict the orbital motion to lie in the horizontal plane, so $z=p_z=0$.
Then the electric field is radial
and the problem reduces to relativistic charged particle orbital and spin motion in a central field of force.
It is standard to treat central force problems using cylindrical polar coordinates $(r,\theta,z)$.
I define the dimensionless variable $\xi=x/r_0$, so $r = r_0(1+\xi)$.
I denote the electric field index by $n$.
The field and potential in the horizontal plane are given by
\begin{subequations}
\label{eq:fieldpot}
\begin{align}
\bm{E} &= -E_0\,\frac{r_0^{1+n}}{r^{1+n}}\,\hat{\bm{r}} \,,
\\
V(r) &= \begin{cases} \displaystyle \frac{E_0 r_0}{n}\,\Bigl(1 - \frac{r_0^n}{r^n}\Bigr) &\qquad (n>0)
\\
\displaystyle E_0 r_0\,\ln\frac{r}{r_0}\phantom{\Biggl|}  &\qquad (n=0) \,.
\end{cases}
\end{align}
\end{subequations}
The potential is normalized so that $V=0$ at $r=r_0$.
Note that, although  in principle the analytical formulas derived below are valid for all $n>0$,
in practice I treated only values $0\le n \le1$ in all of my numerical computations below.
This is because it is well known that for a homogeneous weak-focusing all-electric ring, 
the horizontal betatron tune is given by \cite{ManeElectrostaticBending2012,Laslett_elec_ring}
$\nu_x = \sqrt{2-n-\beta_0^2}$,
and this can possibly be imaginary if $n>1$, depending on the value of $\beta_0$.
Next, the centripetal force yields
\bq
\label{eq:centripetal}
\frac{\gamma m v_\theta^2}{r} = eE_0\,\frac{r_0^{1+n}}{r^{1+n}}  \,.
\eq
Hence $\gamma mv_\theta^2 = eE_0r_0(r_0/r)^n$.
On the reference orbit, $\gamma=\gamma_0$ and $v_\theta=v_0$, hence
$eE_0r_0 = p_0v_0 = mc^2 \gamma_0\beta_0^2$.

Note that if the orbital motion lies in the horizontal plane, then the spin precession vector is vertical:
$\bm{\Omega} \parallel \bm{\beta}\times\bm{r} \parallel \hat{\bm{z}}$.
Hence the spin precesses around the vertical axis, 
i.e.~the vertical component of the spin is a constant of the motion.
Hence, to analyze the spin decoherence rate, it is only the horizontal spin components of the particles which decohere.
Hence I restrict the spin to also lie in the horizontal plane in the analysis below; no generality is lost by doing so.
Furthermore, the initial condition in an EDM experiment is that the spins are injected along the reference axis 
$\bm{s} = \hat{\bm{\theta}}$, which lies in the horizontal plane, by definition.
I next define the angle $\alpha$,
which is the angle between the spin unit vector $\bm{s}$ and
the unit vector in the direction of the velocity $\hat{\bm{\beta}}$.
I follow the definition in \cite{TalmanIPAC2012},
in which we go counterclocksise from $\hat{\bm{\beta}}$ to $\bm{s}$.
Then
\bq
\bm{s}\cdot\hat{\bm{\beta}} = \cos\alpha \,,\qquad
\bm{s}\times\hat{\bm{\beta}} = -\sin\alpha\,\hat{\bm{z}} \,.
\eq
Then, using eq.~\eqref{eq:helicityefield} and $v_\theta = c\beta_\theta = r\dot{\theta}$,
\bq
\begin{split}
\frac{d\alpha}{dt} &= \frac{e}{mc}\biggl(a - \frac{1}{\beta^2\gamma^2}\biggr) (\bm{\beta}\times\bm{E})\cdot\hat{\bm{z}}
\\
&= -\frac{e}{mc^2}\biggl(a - \frac{1}{\beta^2\gamma^2}\biggr) r\dot{\theta}\,
\frac{E_0r_0^{1+n}}{r^{1+n}} (\hat{\bm{\theta}}\times\hat{\bm{r}})\cdot\hat{\bm{z}}
\\
&= \frac{eE_0r_0}{mc^2}\biggl(a - \frac{1}{\beta^2\gamma^2}\biggr)\,\frac{r_0^n}{r^n}\, \dot{\theta}\,.
\end{split}
\eq
Hence, using $eE_0r_0 = mc^2 \gamma_0\beta_0^2$,
\bq
\label{eq:dadtheta}
\frac{d\alpha}{d\theta} = \gamma_0\beta_0^2\, \biggl(a - \frac{1}{\beta^2\gamma^2}\biggr)\,\frac{r_0^n}{r^n} \,.
\eq
This is applicable for any value of the reference momentum
but the case of interest below is when the reference is the magic gamma
\bq
\label{eq:dadthetamagicgamma}
\frac{d\alpha}{d\theta} = \frac{1}{\gamma_0}\Bigl(1 - \frac{\beta_0^2\gamma_0^2}{\beta^2\gamma^2}\Bigr)\,\frac{r_0^n}{r^n} 
= \frac{1}{\gamma_0}\Bigl(1 - \frac{p_0^2}{p^2}\Bigr)\,\frac{r_0^n}{r^n} \,.
\eq

Let us make some general observations using this expression.
On the reference orbit, the term in parentheses vanishes, by definition of the magic gamma.
If we define `$\epsilon$' as a generic `small parameter' for an off-axis orbit
(for example the amplitude of a betatron oscillation or an energy offset)
then we can write
\bq
1 - \frac{\beta_0^2\gamma_0^2}{\beta^2\gamma^2} = \mathcal{C}_1\,\epsilon + \mathcal{C}_2\,\epsilon^2 + \cdots \,.
\eq
Here $\mathcal{C}_1$ and $\mathcal{C}_2$ are quantities which do not depend on $\epsilon$, and their internal details do not matter.
The other terms outside the parentheses are of $O(1)$ on the reference orbit, 
say $\mathcal{D}_0(1 + \mathcal{D}_1\,\epsilon + \cdots)$
(here $\mathcal{D}_0$ and $\mathcal{D}_1$ are also quantities which do not depend on $\epsilon$).
Then we may write, overall
\bq
\begin{split}
\frac{d\alpha}{d\theta} &= \mathcal{D}_0(1 + \mathcal{D}_1\,\epsilon +\cdots)(\mathcal{C}_1\,\epsilon + \mathcal{C}_2\,\epsilon^2 +\cdots) 
\\
&\propto \mathcal{C}_1\,\epsilon + (\mathcal{C}_2 + \mathcal{C}_1\mathcal{D}_1)\,\epsilon^2  + O(\epsilon^3) \,.
\end{split}
\eq
Hence we require the term in parentheses to $O(\epsilon^2)$, as expected, 
but we {\em also} require the other terms to $O(\epsilon)$.
We {\em cannot} approximate the outside terms as $O(1)$ constants;
to do so loses the contribution of the term in $\mathcal{C}_1\mathcal{D}_1$.

Returning to the main thread, let us write $\gamma=\gamma_0 +\Delta\gamma$ to obtain
\bq
\begin{split}
1 - \frac{\beta_0^2\gamma_0^2}{\beta^2\gamma^2} &=
1 - \frac{\beta_0^2\gamma_0^2}{\gamma^2-1} 
\\
&= 1 - \frac{\beta_0^2\gamma_0^2}{\gamma_0^2(1+\Delta\gamma/\gamma_0)^2-1} 
\\
&= 1 - \frac{\beta_0^2}{\beta_0^2+2(\Delta\gamma/\gamma_0) + (\Delta\gamma/\gamma_0)^2} 
\\
&= 1 - \frac{1}{1+(2/\beta_0^2)(\Delta\gamma/\gamma_0) + (1/\beta_0^2)(\Delta\gamma/\gamma_0)^2} 
\\
&\simeq \frac{2}{\beta_0^2}\,\frac{\Delta\gamma}{\gamma_0}
+\frac{1}{\beta_0^2}\,\Bigl(\frac{\Delta\gamma}{\gamma_0}\Bigr)^2
-\frac{4}{\beta_0^4}\,\Bigl(\frac{\Delta\gamma}{\gamma_0}\Bigr)^2 \,.
\end{split}
\eq
Next we expand $r = r_0(1+\xi)$ to obtain
\bq
\frac{r_0^n}{r^n} = \frac{1}{(1+\xi)^n} \simeq 1 - n\xi + \frac{n(n+1)}{2}\,\xi^2 \,.
\eq
Then, noting that both $\Delta\gamma/\gamma_0$ and $\xi$ are of the first order in small quantities
\bq
\label{eq:dadtseries}
\begin{split}
\frac{d\alpha}{d\theta} &\simeq 
\frac{1}{\gamma_0}\,\biggl[\,
\frac{2}{\beta_0^2}\,\frac{\Delta\gamma}{\gamma_0}
+\frac{\beta_0^2-4}{\beta_0^4}\,\Bigl(\frac{\Delta\gamma}{\gamma_0}\Bigr)^2 
\biggr]\,
(1 - n\xi)
\\
&\simeq 
\frac{2}{\gamma_0\beta_0^2}\,\frac{\Delta\gamma}{\gamma_0}
+\frac{\beta_0^2-4}{\gamma_0\beta_0^4}\,\Bigl(\frac{\Delta\gamma}{\gamma_0}\Bigr)^2 
-\frac{2n}{\gamma_0\beta_0^2}\,\frac{\Delta\gamma}{\gamma_0}\,\xi \,.
\end{split}
\eq
Next, we decompose the motion into betatron oscillations and energy offset terms.
I do not say `synchrotron oscillations' because $\Delta H/H_0$ is a constant of the motion in this model.
Using the subscripts `$\beta$' to denote `betatron oscillation' and `$d$' for `dispersion,' we write
\bq
\xi = \xi_\beta + \xi_d \,,\qquad
\Delta\gamma = \Delta\gamma_\beta + \Delta\gamma_d \,.
\eq
(Note that $\Delta\gamma_\beta=0$ in an all-magnetic ring.)
Because the terms in the betatron motion and the energy offsets are statistically independent variables,
all averages over cross terms with `$\beta$' and `$d$' subscripts will vanish.
Hence we can group the terms into separate contributions from the betatron motion and the energy offset
\begin{align}
\label{eq:dadtbetadispavg}
\biggl\langle\frac{d\alpha}{d\theta}\biggr\rangle &=
\biggl\langle\frac{d\alpha_\beta}{d\theta}\biggr\rangle
+ \biggl\langle\frac{d\alpha_d}{d\theta}\biggr\rangle \,,
\\
\label{eq:dadtbetaavg}
\frac{d\alpha_\beta}{d\theta} &\simeq
\frac{2}{\gamma_0\beta_0^2}\,\frac{\Delta\gamma_\beta}{\gamma_0}
+\frac{\beta_0^2-4}{\gamma_0\beta_0^4}\,
\Bigl(\frac{\Delta\gamma_\beta}{\gamma_0}\Bigr)^2
-\frac{2n}{\gamma_0\beta_0^2}\,
\frac{\Delta\gamma_\beta}{\gamma_0}\,\xi_\beta \,,
\\
\label{eq:dadtdispavg}
\frac{d\alpha_d}{d\theta} &\simeq
\frac{2}{\gamma_0\beta_0^2}\,\frac{\Delta\gamma_d}{\gamma_0}
+\frac{\beta_0^2-4}{\gamma_0\beta_0^4}\,
\Bigl(\frac{\Delta\gamma_d}{\gamma_0}\Bigr)^2
-\frac{2n}{\gamma_0\beta_0^2}\,
\frac{\Delta\gamma_d}{\gamma_0}\,\xi_d \,.
\end{align}
I therefore determine $\langle d\alpha/d\theta\rangle_\beta$
and $\langle d\alpha/d\theta\rangle_d$ in separate calculations below (in different sections).

\section{\label{sec:offenergy} Off-energy orbits}
\subsection{Spin decoherence}
This problem can be solved exactly.
In this model, the off-energy (dispersion) orbits are circles. 
Following \cite{ManeElectrostaticBending2012},
I define the parameter $\lambda_p = (1/\beta_0^2)(\Delta H/H_0)$,
hence the energy is $H = H_0(1 + \Delta H/H_0) = H_0(1 + \beta_0^2\lambda_p)$.
For simplicity of the notation, 
$\xi$ and $\gamma$ will denote $\xi_d$ and $\gamma_d$, respectively,
in the calculations below.
The centripetal force yields (see eq.~\eqref{eq:centripetal})
\bq
\label{eq:centripetaloffenergy}
\gamma\beta^2 = \gamma - \frac{1}{\gamma} = \gamma_0\beta_0^2\,\frac{r_0^n}{r^n} \,.
\eq
We know that for $n=0$ (logarithmic potential) $\gamma\beta^2 = \gamma_0\beta_0^2$, 
and the solution is $\gamma=\gamma_0$ independent of the radius.
The kinetic energy is the same for all the off-energy dispersion orbits.
Then from eq.~\eqref{eq:dadthetamagicgamma} we have $d\alpha_d/d\theta=0$.
This is the exact solution for off-energy motion in this model for $n=0$.
The orbit radius $r_d$ is fixed by the value of the potential energy as follows
\begin{align}
\frac{H}{mc^2} = \gamma_0(1 + \beta_0^2\lambda_p) &= \gamma_0 + \gamma_0\beta_0^2\,\ln\frac{r_d}{r_0} \,,
\\
\gamma_0\beta_0^2\lambda_p &= \gamma_0\beta_0^2\,\ln\frac{r_d}{r_0} \,,
\\
r_d &= r_0 \,e^{\lambda_p} \,.
\end{align}
For $n>0$ we proceed as follows
(where we use eq.~\eqref{eq:centripetaloffenergy} to replace $r_0^n/r^n$ in terms of $\gamma$)
\begin{align}
\frac{H}{mc^2} &= \gamma + \frac{\gamma_0\beta_0^2}{n}\,\Bigl(1 - \frac{r_0^n}{r^n}\Bigr) \,,
\\
\label{eq:secondline}
\frac{nH}{mc^2} &= n\gamma + \gamma_0\beta_0^2 - \Bigl(\gamma - \frac{1}{\gamma}\Bigr) \,,
\\
0 &= (1-n)\gamma^2 - \Bigl(\gamma_0\beta_0^2 - \frac{nH}{mc^2}\Bigr)\,\gamma - 1 \,.
\end{align}
We want the solution of the quadratic equation for which $\gamma_d>1$. 
Hence
\bq
\label{eq:gammadquadratic}
\gamma_d = \frac{1}{2(1-n)}\,\biggl[\, \Bigl(\gamma_0\beta_0^2 - \frac{nH}{mc^2}\Bigr)
+ \biggl| \sqrt{\Bigl(\gamma_0\beta_0^2 - \frac{nH}{mc^2}\Bigr)^2 + 4(1-n)} \biggr| \,\biggr] \,.
\eq
However, the above yields a $0/0$ expression for $n=1$.
For $n=1$, note from eq.~\eqref{eq:secondline} that
\begin{align}
\frac{H}{mc^2} &= \gamma_d + \gamma_0\beta_0^2 - \Bigl(\gamma_d - \frac{1}{\gamma_d}\Bigr) \,,
\\
\label{eq:invgammadkepler}
\frac{1}{\gamma_d} &= \frac{H}{mc^2} - \gamma_0\beta_0^2  
= \gamma_0(1 - \beta_0^2) + \gamma_0\beta_0^2\lambda_p   
= \frac{1 + \gamma_0^2\beta_0^2\lambda_p}{\gamma_0} \,,
\\
\label{eq:gammadkepler}
\gamma_d &= \frac{\gamma_0}{1 +\gamma_0^2\beta_0^2\lambda_p} \,. 
\end{align}
Then for all $n>0$, we have using eq.~\eqref{eq:centripetaloffenergy} 
\bq
\label{eq:rdoffenergy}
r_d = r_0 \, \biggl[\,\frac{\gamma_0\beta_0^2}{\gamma_d-1/\gamma_d}\,\biggr]^{1/n} \,.
\eq
This expression was used as an initial condition in tracking studies to be reported below.
For the spin precession, from eq.~\eqref{eq:dadthetamagicgamma} and using eq.~\eqref{eq:rdoffenergy}
(here $\beta_d=\sqrt{1-1/\gamma_d^2}$ is defined in the obvious way)
\bq
\begin{split}
\label{eq:dadthetaoffenergy}
\frac{d\alpha_d}{d\theta} 
&= \frac{1}{\gamma_0}\,\Bigl(1 - \frac{\beta_0^2\gamma_0^2}{\beta_d^2\gamma_d^2}\Bigr)\,\frac{r_0^n}{r_d^n}
\\
&= \frac{1}{\beta_0^2\gamma_0^2}\,\Bigl(1 - \frac{\gamma_0^2-1}{\gamma_d^2-1}\Bigr)\,\Bigl(\gamma_d - \frac{1}{\gamma_d}\Bigr)
\\
&= \frac{1}{\beta_0^2\gamma_0^2}\,\frac{\gamma_d^2-\gamma_0^2}{\gamma_d}
\\
&= \frac{1}{\gamma_0\beta_0^2}\,\Bigl(\frac{\gamma_d}{\gamma_0} - \frac{\gamma_0}{\gamma_d}\Bigr) \,.
\end{split}
\eq
This works also for $n=0$ because $\gamma_d=\gamma_0$ so the right-hand side vanishes.
Hence this is the exact solution for all $n \ge 0$.

Nevertheless, the above solution is not easily visualizable 
as a function of $\lambda_d$ for small $|\lambda_d|$.
For $n=1$, we can derive the exact solution 
using eqs.~\eqref{eq:invgammadkepler} and \eqref{eq:gammadkepler}
\bq
\begin{split}
\frac{d\alpha_d}{d\theta} &= \frac{1}{\gamma_0\beta_0^2}\,\biggl[\,
\frac{1}{1 +\gamma_0^2\beta_0^2\lambda_p} - (1 + \gamma_0^2\beta_0^2\lambda_p) \,\biggr]
\\
&= \frac{1}{\gamma_0\beta_0^2}\,
\frac{1 - (1 +\gamma_0^2\beta_0^2\lambda_p)^2}{1 +\gamma_0^2\beta_0^2\lambda_p}
\\
&= -\gamma_0\lambda_p\,\frac{2 + \gamma_0^2\beta_0^2\lambda_p}{1 +\gamma_0^2\beta_0^2\lambda_p} \,.
\end{split}
\eq
For $n\ne0$ and $n\ne1$ we expand in powers of $\lambda_p$.
We set $\gamma_d=\gamma_0(1+\Delta\gamma_d/\gamma_0)$.
We employ eq.~\eqref{eq:secondline} which we write in the form
\bq
\begin{split}
0 &= (1-n)\gamma_d -\gamma_0\beta_0^2 +\frac{nH}{mc^2} -\frac{1}{\gamma_d}
\\
&= (1-n)\gamma_0\Bigl(1+\frac{\Delta\gamma_d}{\gamma_0}\Bigr) 
-\gamma_0\beta_0^2 +n\gamma_0(1+\beta_0^2\lambda_p) 
-\frac{1}{\gamma_0}\Bigl(1+\frac{\Delta\gamma_d}{\gamma_0}\Bigr)^{-1}
\\
&\simeq (1-n)\gamma_0^2\Bigl(1+\frac{\Delta\gamma_d}{\gamma_0}\Bigr) 
-\gamma_0^2\beta_0^2 +n\gamma_0^2(1+\beta_0^2\lambda_p) 
-\biggl[\,1-\frac{\Delta\gamma_d}{\gamma_0} +\Bigl(\frac{\Delta\gamma_d}{\gamma_0}\Bigr)^2\,\biggr]
\\
&= (1-n)\gamma_0^2 -\gamma_0^2\beta_0^2 +n\gamma_0^2(1+\beta_0^2\lambda_p) -1
+(1+(1-n)\gamma_0^2)\,\frac{\Delta\gamma_d}{\gamma_0}
- \Bigl(\frac{\Delta\gamma_d}{\gamma_0}\Bigr)^2 
\\
&= n\beta_0^2\gamma_0^2\lambda_p 
+(1 + (1-n)\gamma_0^2)\,\frac{\Delta\gamma_d}{\gamma_0}
- \Bigl(\frac{\Delta\gamma_d}{\gamma_0}\Bigr)^2 \,.
\end{split}
\eq
We write this in the form
\bq
\frac{\Delta\gamma_d}{\gamma_0} = -\frac{n\beta_0^2\gamma_0^2}{1 + (1-n)\gamma_0^2}\,\lambda_p
+\frac{1}{1 + (1-n)\gamma_0^2}\Bigl(\frac{\Delta\gamma_d}{\gamma_0}\Bigr)^2 \,.
\eq
Then approximately, to the first order,
\bq
\label{eq:dgoffenergyfirstorder} 
\frac{\Delta\gamma_d}{\gamma_0} \simeq -\frac{n\beta_0^2\gamma_0^2}{1+(1-n)\gamma_0^2}\,\lambda_p \,.
\eq
Next, to the second order,
\bq
\label{eq:dgoffenergy}
\frac{\Delta\gamma_d}{\gamma_0} \simeq -\frac{n\beta_0^2\gamma_0^2}{1+(1-n)\gamma_0^2}\,\lambda_p 
+ \frac{n^2\beta_0^4\gamma_0^4}{(1+(1-n)\gamma_0^2)^3}\,\lambda_p^2 \,.
\eq
Then for the spin precession we obtain, 
using eqs.~\eqref{eq:dadthetaoffenergy} and \eqref{eq:dgoffenergy},
\bq
\begin{split}
\frac{d\alpha_d}{d\theta} 
&= \frac{1}{\gamma_0\beta_0^2}\,\Bigl(\frac{\gamma_d}{\gamma_0} - \frac{\gamma_0}{\gamma_d}\Bigr) 
\\
&= \frac{1}{\gamma_0\beta_0^2}\,\biggl[\, 1 + \frac{\Delta\gamma_d}{\gamma_0}
- \Bigl(1 + \frac{\Delta\gamma_d}{\gamma_0}\Bigr)^{-1} \,\biggr]
\\
&\simeq \frac{1}{\gamma_0\beta_0^2}\,\biggl[\, 2\,\frac{\Delta\gamma_d}{\gamma_0}
- \Bigl(\frac{\Delta\gamma_d}{\gamma_0}\Bigr)^2 \,\biggr]
\\
&\simeq 
-\frac{2n\gamma_0}{1+(1-n)\gamma_0^2}\,\lambda_p 
+ \frac{2n^2\beta_0^2\gamma_0^3}{(1+(1-n)\gamma_0^2)^3}\,\lambda_p^2 
-\frac{n^2\beta_0^2\gamma_0^3}{(1+(1-n)\gamma_0^2)^2}\,\lambda_p^2 
\\
&= 
-\frac{2n\gamma_0}{1+(1-n)\gamma_0^2}\,\lambda_p 
+ \frac{n^2\beta_0^2\gamma_0^3(1-(1-n)\gamma_0^2)}{(1+(1-n)\gamma_0^2)^3}\,\lambda_p^2 \,.
\end{split}
\eq
Hence $d\alpha_d/d\theta=0$ for $n=0$, as we know from the exact solution.
Since the energy $H$ is a dynamical invariant for this model
(hence also $\Delta H/H_0$ and $\lambda_p$)
we can leave the results in the above form.
If we average over a statistical distribution of energies, we obtain
\bq
\label{eq:dadthetaoffenergyavg}
\biggl\langle \frac{d\alpha_d}{d\theta} \biggr\rangle
\simeq
-\frac{2n\gamma_0}{1+(1-n)\gamma_0^2}\,\langle \lambda_p \rangle
+\frac{n^2\beta_0^2\gamma_0^3(1-(1-n)\gamma_0^2)}{(1+(1-n)\gamma_0^2)^3}\,\langle \lambda_p^2 \rangle \,.
\eq

\subsection{Orbit radius}
For the record, let us calculate the orbit radius $r_d$ perturbatively.
To save tedious algebra, I calculate to $O(\lambda_p)$ only.
Using eq.~\eqref{eq:dgoffenergyfirstorder} in eq.~\eqref{eq:rdoffenergy},
\bq
\begin{split}
\frac{r_d}{r_0} &= (\gamma_0\beta_0^2)^{1/n}\,\biggl[\,\gamma_d-\frac{1}{\gamma_d}\,\biggr]^{-1/n} 
\\
&\simeq (\gamma_0\beta_0^2)^{1/n}\,\biggl[\,\gamma_0\Bigl(1+\frac{\Delta\gamma_d}{\gamma_0}\Bigr) 
-\frac{1}{\gamma_0}\Bigl(1-\frac{\Delta\gamma_d}{\gamma_0}\Bigr) \,\biggr]^{-1/n} 
\\
&= (\gamma_0\beta_0^2)^{1/n}\,\biggl[\,\gamma_0\beta_0^2 + \frac{\gamma_0^2+1}{\gamma_0}\,\frac{\Delta\gamma_d}{\gamma_0} \,\biggr]^{-1/n} 
\\
&= \biggl[\,1 + \frac{\gamma_0^2+1}{\beta_0^2\gamma_0^2}\,\frac{\Delta\gamma_d}{\gamma_0} \,\biggr]^{-1/n} 
\\
&\simeq 1 - \frac{1}{n}\frac{\gamma_0^2+1}{\beta_0^2\gamma_0^2}\,\frac{\Delta\gamma_d}{\gamma_0}
\\
&\simeq 1 + \frac{1}{n}\,\frac{\gamma_0^2+1}{\beta_0^2\gamma_0^2}\,\frac{n\beta_0^2\gamma_0^2}{1+(1-n)\gamma_0^2}\,\lambda_p 
\\
&= 1 + \frac{\gamma_0^2+1}{1+(1-n)\gamma_0^2}\,\lambda_p \,.
\end{split}
\eq

\subsection{Synchrotron oscillations}
I conclude this section with a brief commentary on synchrotron oscillations in this model.
{\bf{\em \color{red}
Note very carefully that synchrotron oscillations are variations of the 
{\em total energy}, i.e.~$\Delta H/H_0$ and {\em not} $\Delta\gamma/\gamma_0$.
An rf cavity changes the \underline{value of $H$}, which may or may not change the value of $\gamma$.}}
When electrostatic guiding and focusing fields are present, 
the value of $\gamma$ may in fact {\em not} vary at all in a synchrotron oscillation.
In particular, it is well known (and we have seen above) that for $n=0$ (logarithmic potential),
$\gamma_d = \gamma_0$ on all the dispersion orbits, and does {\em not} depend on the total energy $H$.
Hence we see that 
$\Delta\gamma_d/\gamma_0 = 0$ for synchrotron oscillations in a logarithmic potential.

{\em N.B.: In this model, the dispersion is uniform around the circumference, hence there is synchrobetatron coupling.
Hence to justify the above statements rigorously,
we require the synchrotron tune to be very small, i.e.~$\nu_s \ll 1$,
to minimize the synchrobetatron coupling.
The quasistatic approximation is also required to justify that we can continue to express the
motion in $x$ and $p_x$ as a sum of betatron oscillations and dispersion orbits,
instead of a general fully coupled symplectic formalism.
}

\section{\label{sec:betatronoscs} Betatron oscillations}
\subsection{Spin decoherence rate}
Next I treat the betatron oscillations.
I fix $H=H_0=\gamma_0 mc^2$ in the calculations below.
For simplicity of the notation, 
$\xi$ and $\gamma$ will denote $\xi_\beta$ and $\gamma_\beta$, respectively,
in the calculations below.
We must expand $r$ and $\gamma$ for the off-axis betatron motion.
Note that 
\bq
\label{eq:phioffaxis}
\Phi = \frac{mc^2\gamma_0\beta_0^2}{n}\,\Bigl(1 - \frac{1}{(1+\xi)^n}\Bigr) 
\simeq mc^2\gamma_0\beta_0^2 \Bigl(\xi - \frac{(1+n)}{2}\,\xi^2\Bigr) \,.
\eq
The final expression is applicable also for $n=0$ (logarithmic potential).
Note also that $\gamma = H/(mc^2) - \Phi/(mc^2)$
and as stated above, I fix $H = H_0$.
Then, using eq.~\eqref{eq:phioffaxis},
\bq
\label{eq:gamma}
\frac{\gamma}{\gamma_0} = 1 - \frac{\Phi}{\gamma_0 mc^2}
\simeq 1 - \beta_0^2\Bigl(\xi - \frac{1+n}{2}\,\xi^2\Bigr)  \,.
\eq
Then
\bq
\frac{\Delta\gamma}{\gamma_0} \simeq - \beta_0^2\Bigl(\xi - \frac{1+n}{2}\,\xi^2\Bigr)  \,.
\eq
Substituting in eq.~\eqref{eq:dadtbetaavg} yields
\bq
\label{eq:dadt_taylor}
\begin{split}
\frac{d\alpha_\beta}{d\theta} &\simeq
-\frac{2}{\gamma_0}\,\Bigl(\xi - \frac{1+n}{2}\,\xi^2\Bigr)
+\frac{\beta_0^2-4}{\gamma_0}\,
\Bigl(\xi - \frac{1+n}{2}\,\xi^2\Bigr)^2
+\frac{2n}{\gamma_0}\,\Bigl(\xi - \frac{1+n}{2}\,\xi^2\Bigr) \xi
\\
&\simeq -\frac{1}{\gamma_0}\,\bigr[\, 2\xi -(1+n)\xi^2 - (\beta_0^2-4)\xi^2 - 2n\xi^2 \,\bigr]
\\
&= -\frac{1}{\gamma_0}\,\bigr[\, 2\xi +(3 -3n -\beta_0^2)\xi^2 \,\bigr] \,.
\end{split}
\eq
To average over orbits, we need to derive expressions for $\langle \xi\rangle$ and $\langle \xi^2\rangle$.
This will be done below using canonical transformations to diagonalize the Hamiltonian.

\subsection{Canonical transformations}
Henceforth I set $c=1$.
I shall work with $\xi=x/r_0$ below.
To preserve the Hamiltonian structure of the equations, 
we scale the independent variable to $\theta=s/r_0$:
\bq
\frac{d\xi}{d\theta} = \frac{dx}{ds} = \frac{\partial K}{\partial p_x} \,,\qquad
\frac{dp_x}{d\theta} = r_0\,\frac{dp_x}{ds} = -\frac{\partial K}{\partial \xi} \,.
\eq
Next, we scale the momentum $p_x = p_0\,p_\xi$.
To preserve the Hamiltonian structure of the equations, we divide $K$ by $p_0$:
\bq
\frac{d\xi}{d\theta} = \frac{\partial (K/p_0)}{\partial p_\xi} \,,\qquad
\frac{dp_\xi}{d\theta} = -\frac{\partial (K/p_0)}{\partial \xi} \,.
\eq
Hence we define $\bar{K}=K/p_0$
\bq
\begin{split}
\bar{K} = \frac{K}{p_0} = -\frac{p_s}{p_0} &= 
-\frac{1+\xi}{p_0}\,\biggl[\, m^2\gamma_0^2\Bigl(1 -\frac{\Phi}{m\gamma_0}\Bigr)^2 - m^2 - p_0^2p_\xi^2 \,\biggr]^{\frac12} 
\\
&= -(1+\xi)\,\biggl[\, 1 -\frac{2m\gamma_0\Phi}{p_0^2} + \frac{\Phi^2}{p_0^2} - p_\xi^2 \,\biggr]^{\frac12} \,.
\end{split}
\eq
Define $\kappa = p_s/p_0$.
Note that $K$ is invariant along an orbit, hence $p_s$ and hence $\kappa$ are also invariant. Hence $p_s=p_{s0}$ and $\kappa=\kappa_0$.
Note that the value of $\kappa$ must be precomputed using the initial data.
We can employ the equivalent Hamiltonian
\bq
K_1 = \frac{(1+\xi)^2}{2\kappa}\,\biggl[\, p_\xi^2 -1 +\frac{2m\gamma_0\Phi}{p_0^2} - \frac{\Phi^2}{p_0^2} \,\biggr] \,.
\eq
The partial derivatives of all dynamical variables are the same using $K_1$ and $\bar{K}$.
We work with $K_1$ below. 
I shall treat only the case $H = \gamma_0 m$ below.
Then there is no off-energy dispersion orbit:
the orbital motion in $(x,p_x)$, or $(\xi,p_\xi)$, consists purely of betatron oscillations.
I expand $K_1$ in a Taylor series in powers of $\xi$, up to the fourth power.
The first step is to expand the potential in a Taylor series to the fourth power in $\xi$
(note that the Taylor series works also for a logarithmic potential, i.e.~$n=0$)
\bq
\Phi \simeq eE_0r_0\,\biggl[\, \xi-\frac{(1+n)\xi^2}{2!}+\frac{(1+n)(2+n)\xi^3}{3!}-\frac{(1+n)(2+n)(3+n)\xi^4}{4!} \,\biggr] \,.
\eq
Define
\bq
K_a \equiv \frac{2m\gamma_0\Phi}{p_0^2}
\simeq 2\,\biggl[\, \xi-\frac{(1+n)\xi^2}{2!}+\frac{(1+n)(2+n)\xi^3}{3!}-\frac{(1+n)(2+n)(3+n)\xi^4}{4!} \,\biggr] \,.
\eq
Then
\bq
\begin{split}
&\quad (1+\xi)^2K_a 
\\
&\simeq
2(1+2\xi+\xi^2)\,\biggl[\, \xi-\frac{(1+n)\xi^2}{2!}+\frac{(1+n)(2+n)\xi^3}{3!}-\frac{(1+n)(2+n)(3+n)\xi^4}{4!} \,\biggr] 
\\
&\simeq
2\,\biggl[\, \xi-\frac{(1+n)\xi^2}{2!}+\frac{(1+n)(2+n)\xi^3}{3!}-\frac{(1+n)(2+n)(3+n)\xi^4}{4!} \,\biggr] 
\\
&\quad
+2\,\biggl[\, 2\xi^2-(1+n)\xi^3+\frac{(1+n)(2+n)\xi^4}{3} \,\biggr] 
+2\,\biggl[\, \xi^3-\frac{(1+n)\xi^4}{2} \,\biggr] 
\\
&\simeq 
2\,\biggl[\, \xi+\frac{(3-n)\xi^2}{2}+\frac{(1-n)(2-n)\xi^3}{6}+\frac{[4(1+n)(1+2n) -(1+n)(2+n)(3+n)]\xi^4}{24} \,\biggr] 
\\
&\simeq 
2\,\biggl[\, \xi+\frac{(3-n)\xi^2}{2}+\frac{(1-n)(2-n)\xi^3}{6}-\frac{(1+n)(2-3n+n^2)\xi^4}{24} \,\biggr] 
\\
&\simeq 
2\,\biggl[\, \xi+\frac{(3-n)\xi^2}{2}+\frac{(1-n)(2-n)\xi^3}{6}-\frac{(1+n)(1-n)(2-n)\xi^4}{24} \,\biggr] \,.
\end{split}
\eq
Next define $K_b = \Phi^2/p_0^2$.
Then
\bq
\begin{split}
\frac{K_b}{\beta_0^2} &\simeq
\biggl[\, \xi-\frac{(1+n)\xi^2}{2!}+\frac{(1+n)(2+n)\xi^3}{3!}-\frac{(1+n)(2+n)(3+n)\xi^4}{4!} \,\biggr]^2
\\
&\simeq
\xi^2 - (1+n)\xi^3 + \frac{(1+n)^2\xi^4}{4} +\frac{(1+n)(2+n)\xi^4}{3}
\\
&= \xi^2 - (1+n)\xi^3 + \frac{(1+n)(3+3n+8+4n)\xi^4}{12} 
\\
&= \xi^2 - (1+n)\xi^3 + \frac{(1+n)(11+7n)\xi^4}{12} \,.
\end{split}
\eq
Then
\bq
\begin{split}
(1+\xi)^2\frac{K_b}{\beta_0^2} &\simeq (1+2\xi+\xi^2)\,\biggl[\, \xi^2 - (1+n)\xi^3 + \frac{(1+n)(11+7n)\xi^4}{12} \,\biggr]
\\
&\simeq \xi^2 - (1+n)\xi^3 + \frac{(1+n)(11+7n)\xi^4}{12} 
+ 2\xi^3 - 2(1+n)\xi^4 + \xi^4
\\
&\simeq \xi^2 + (1-n)\xi^3 + \frac{11+18n+7n^2 -24 - 48n +12}{12}\,\xi^4
\\
&\simeq \xi^2 + (1-n)\xi^3 - \frac{1+30n-7n^2}{12}\,\xi^4 \,.
\end{split}
\eq
Then
\bq
\begin{split}
K_1 &\simeq \frac{1}{2\kappa} \,
\biggl[\, p_\xi^2(1+2\xi+\xi^2) -1 -2\xi -\xi^2
\\
&\qquad\qquad
+2\,\biggl( \xi+\frac{(3-n)\xi^2}{2}+\frac{(1-n)(2-n)\xi^3}{6}-\frac{(1+n)(1-n)(2-n)\xi^4}{24} \,\biggr) 
\\
&\qquad\qquad
-\beta_0^2\,\biggl( \xi^2 + (1-n)\xi^3 - \frac{1+30n-7n^2}{12}\,\xi^4 \biggr) \,\biggr]
\\
&= \textrm{(const.)} 
+\frac{p_\xi^2}{2\kappa} + \frac{\kappa}{2}\,\frac{2-n-\beta_0^2}{\kappa^2}\,\xi^2
\\
&\qquad\qquad\quad
+\frac{1}{2\kappa} \,
\biggl[\, p_\xi^2(2\xi+\xi^2) +(1-n)\biggl(\frac{2-n}{3} -\beta_0^2\biggr)\xi^3
\\
&\qquad\qquad\qquad\qquad
-\frac{1}{12}\biggl( (1+n)(1-n)(2-n) -\beta_0^2\,(1+30n-7n^2) \biggr) \xi^4
\,\biggr] \,.
\end{split}
\eq
We discard the constant term, and separate $K_1$ into quadratic and anharmonic terms. 
The quadratic terms describe the motion of a particle of mass $\kappa$,
with a tune $\nu_x = \sqrt{2-n-\beta_0^2}/\kappa$. 
We can define action-angle variables $(J,\phi)$ for the linear dynamical motion
\bq
\label{eq:aalin}
\xi = \sqrt{2J/(\kappa\nu_x)}\,\cos\phi \,,\qquad
p_\xi = -\sqrt{2J\kappa\nu_x}\,\sin\phi \,,\qquad 
\frac{d\phi}{d\theta} = \nu_x \,.
\eq
Then the Hamiltonian in (linear dynamical) action-angle variables is
\bq
\begin{split}
\label{eq:kaa}
\mathcal{K} = \nu_x J 
&+\nu_x J\sin^2\phi\,\biggl[\,2\Bigl(\frac{2J}{\kappa\nu_x}\Bigr)^{1/2}\,\cos\phi +\frac{2J}{\kappa\nu_x}\,\cos^2\phi\,\biggl]
\\
&+\frac{1-n}{2\kappa}\Bigl(\frac{2-n}{3}-\beta_0^2\Bigr)
\Bigl(\frac{2J}{\kappa\nu_x}\Bigr)^{3/2}\,\cos^3\phi
\\
&-\frac{1}{6\kappa}\Bigl(\frac{J}{\kappa\nu_x}\Bigr)^2
\biggl( (1+n)(1-n)(2-n) -\beta_0^2\,(1+30n-7n^2) \biggr) \,\cos^4\phi \,.
\end{split} 
\eq
We use the trigonometric identities
\begin{subequations}
\begin{align}
\sin^2\phi\cos\phi &= \frac12\sin(2\phi)\sin\phi 
= \frac14\,\bigl[\, \cos\phi - \cos(3\phi) \,\bigr] \,,
\\
\sin^2\phi\cos^2\phi &= \frac14\sin^2(2\phi)
= \frac18\,\bigl[\, 1 - \cos(4\phi) \,\bigr] \,,
\\
\cos^3\phi &= \frac14\,\bigl[\,3\cos\phi +\cos(3\phi)\,\bigr] \,,
\\
\cos^4\phi &= \frac18\,\bigl[\,3+4\cos(2\phi)+\cos(4\phi) \,\bigr] \,.
\end{align}
\end{subequations}
Then the expansion in Fourier harmonics is
\bq
\begin{split}
\label{eq:kaa1}
\mathcal{K}
= \nu_x J 
&
+\frac{\kappa\nu_x^2}{4}\,\Bigl(\frac{2J}{\kappa\nu_x}\Bigr)^{3/2}\,\bigl[\, \cos\phi - \cos(3\phi) \,\bigr]
\\
& +\frac{J^2}{4\kappa}\,\bigl[\, 1 - \cos(4\phi) \,\bigr]
\\
& +\frac{1-n}{8\kappa}\,\Bigl(\frac{2-n}{3} -\beta_0^2\Bigr)
\Bigl(\frac{2J}{\kappa\nu_x}\Bigr)^{3/2}\,\bigl[\,3\cos\phi +\cos(3\phi)\,\bigr]
\\
& -\frac{J^2}{48\kappa^3\nu_x^2}\,
\biggl[ (1+n)(1-n)(2-n) -\beta_0^2\,(1+30n-7n^2) \biggr] \,
\bigl[\, 3 +4\cos(2\phi) +\cos(4\phi)\,\bigr] 
\\
= \nu_x J 
& +\Bigl(\frac{2J}{\kappa\nu_x}\Bigr)^{3/2}\,\biggl\{ \frac{\kappa\nu_x^2}{4}\,\bigl[\, \cos\phi - \cos(3\phi) \,\bigr]
+\frac{1-n}{8\kappa}\,\Bigl(\frac{2-n}{3} -\beta_0^2\Bigr)
\,\bigl[\,3\cos\phi +\cos(3\phi)\,\bigr] \biggr\}
\\
& +\frac{J^2}{4\kappa}\,\bigl[\, 1 - \cos(4\phi) \,\bigr]
\\
& -\frac{J^2}{48\kappa^3\nu_x^2}\,
\biggl[ (1+n)(1-n)(2-n) -\beta_0^2\,(1+30n-7n^2) \biggr] \,
\bigl[\, 3 +4\cos(2\phi) +\cos(4\phi)\,\bigr] \,.
\end{split}
\eq
We eliminate the terms in $J^{3/2}$, all of which are nonsecular.
Let the new action-angle variables be $(J_1,\phi_1)$.
The generating function is
\bq
\label{eq:gf1}
\begin{split}
\mathcal{G}_1 &= \phi J_1 
\\
&\quad
-\Bigl(\frac{2J_1}{\kappa\nu_x}\Bigr)^{3/2}\,\biggl\{ \frac{\kappa\nu_x}{4}\,\bigl[\, \sin\phi - \frac13\,\sin(3\phi) \,\bigr]
+\frac{1-n}{8\kappa\nu_x}\Bigl(\frac{2-n}{3}-\beta_0^2\Bigr)\,\bigl[\,3\sin\phi +\frac13\,\sin(3\phi) \,\bigr] \biggr\} \,.
\end{split}
\eq
The new angle variable $\phi_1$ is given by
\bq
\begin{split}
\phi_1 &= \frac{\partial\mathcal{G}_1}{\partial J_1}
\\
&= \phi 
-\frac32\Bigl(\frac{2}{\kappa\nu_x}\Bigr)^{3/2}J_1^{1/2}\,\biggl\{ \frac{\kappa\nu_x}{4}\,\bigl[\, \sin\phi - \frac13\,\sin(3\phi) \,\bigr]
\\
&\qquad\qquad\qquad\qquad\qquad\qquad
+\frac{1-n}{8\kappa\nu_x}\Bigl(\frac{2-n}{3}-\beta_0^2\Bigr)\,\bigl[\,3\sin\phi +\frac13\,\sin(3\phi) \,\bigr] \biggr\} \,.
\end{split}
\eq
The old action variable $J$ is given by
\bq
\begin{split}
J &= \frac{\partial\mathcal{G}_1}{\partial\phi}
\\
&= J_1
-\Bigl(\frac{2J_1}{\kappa\nu_x}\Bigr)^{3/2}\,\biggl\{ \frac{\kappa\nu_x}{4}\,\bigl[\, \cos\phi - \cos(3\phi) \,\bigr]
+\frac{1-n}{8\kappa\nu_x}\Bigl(\frac{2-n}{3}-\beta_0^2\Bigr)\,\bigl[\,3\cos\phi +\cos(3\phi) \,\bigr] \biggr\} \,.
\end{split}
\eq
Then we may set $J \simeq J_1$ in the $O(J^2)$ terms in $\mathcal{K}$.
For the $O(J^{3/2})$ terms in $\mathcal{K}$, we may set $\phi \simeq \phi_1$, which yields
\bq
\begin{split}
J^{3/2} &= J_1^{3/2} \,\Bigl( 1 + \frac{\Delta J_1}{J_1}\Bigr)^{3/2} 
\\
&
\simeq  J_1^{3/2} + \frac32 J_1^{1/2} \Delta J_1
\\
&\simeq J_1^{3/2}  
-\frac32\Bigl(\frac{2}{\kappa\nu_x}\Bigr)^{3/2}\,J_1^2\,\biggl\{ \frac{\kappa\nu_x}{4}\,\bigl[\, \cos\phi_1 - \cos(3\phi_1) \,\bigr]
\\
&\qquad\qquad\qquad\qquad\qquad\qquad
+\frac{1-n}{8\kappa\nu_x}\Bigl(\frac{2-n}{3}-\beta_0^2\Bigr)\,\bigl[\,3\cos\phi_1 +\cos(3\phi_1) \,\bigr] \biggr\} \,.
\end{split}
\eq

\noindent
The transformed Hamiltonian is (note that $\nu_x\Delta J_1$ cancels the terms in $J_1^{3/2}$)
\bq
\begin{split}
\mathcal{K}_1 &= \mathcal{K} +\underbrace{\frac{\partial\mathcal{G}_1}{\partial\theta}}_{=0}
\\
&\simeq \nu_x (J_1+\Delta J_1) 
\\
& \quad
+\biggl\{ \frac{\kappa\nu_x^2}{4}\,\bigl[\, \cos\phi - \cos(3\phi) \,\bigr]
+\frac{1-n}{8\kappa}\Bigl(\frac{2-n}{3}-\beta_0^2\Bigr) \,\bigl[\,3\cos\phi +\cos(3\phi) \,\bigr] \biggr\}
\,\Bigl(\frac{2}{\kappa\nu_x}\Bigr)^{\frac32}\,J_1^{\frac32}\Bigl(1+\frac{\Delta J_1}{J_1}\Bigr)^{\frac32}
\\
&\quad +\frac{J_1^2}{4\kappa}\,\bigl[\, 1 - \cos(4\phi_1) \,\bigr]
\\
&\quad -\frac{J_1^2}{48\kappa^3\nu_x^2}\,
\biggl[ (1+n)(1-n)(2-n) -\beta_0^2\,(1+30n-7n^2) \biggr] \,
\bigl[\, 3 +4\cos(2\phi_1) +\cos(4\phi_1)\,\bigr] \,.
\\
&\simeq \nu_x J_1
\\
& \quad
-\frac{12 J_1^2}{\kappa^5\nu_x^4} \,
\biggl[\, \frac{\kappa^2\nu_x^2}{4}\,\bigl[\, \cos\phi_1 - \cos(3\phi_1) \,\bigr]
+\frac{1-n}{8}\Bigl(\frac{2-n}{3}-\beta_0^2\Bigr) \,\bigl[\,3\cos\phi_1 +\cos(3\phi_1) \,\bigr] \,\biggr]^2
\\
&\quad +\frac{J_1^2}{4\kappa}\,\bigl[\, 1 - \cos(4\phi_1) \,\bigr]
\\
&\quad -\frac{J_1^2}{48\kappa^3\nu_x^2}\,
\biggl[ (1+n)(1-n)(2-n) -\beta_0^2\,(1+30n-7n^2) \biggr] \,
\bigl[\, 3 +4\cos(2\phi_1) +\cos(4\phi_1)\,\bigr] 
\\
&\simeq \nu_x J_1 +\frac{J_1^2}{4\kappa}
-\frac{J_1^2}{16\kappa^3\nu_x^2}\,\biggl[ (1+n)(1-n)(2-n) -\beta_0^2\,(1+30n-7n^2) \biggr] 
\\
& \quad
-\frac{3 J_1^2}{8\kappa^5\nu_x^4} \,
\biggl\{
\Bigl[ \kappa^2\nu_x^2 +\frac{1-n}{2}(2-n-3\beta_0^2) \,\Bigr]^2
+\Bigl[ \kappa^2\nu_x^2 -\frac{1-n}{6}(2-n-3\beta_0^2) \Bigr]^2
\,\biggr\}
\\
&\quad +\textrm{(oscillatory)} 
\\
&\simeq \nu_x J_1 +\frac{J_1^2}{4\kappa}
-\frac{J_1^2}{16\kappa^3\nu_x^2}\,\biggl[ (1+n)(1-n)(2-n) -\beta_0^2\,(1+30n-7n^2) \biggr] 
\\
& \quad
-\frac{J_1^2}{8\kappa^5\nu_x^4} \,
\biggl\{ 6\kappa^4\nu_x^4
+2 \kappa^2\nu_x^2 (1-n)(2-n-3\beta_0^2) 
+\frac{5}{6}(1-n)^2(2-n-3\beta_0^2)^2
\,\biggr\}
\\
&\quad +\textrm{(oscillatory)} \,.
\end{split}
\eq
From this we can deduce the leading order tuneshift
\bq
\begin{split}
\nu_{x1} &= \frac{\partial\mathcal{K}_1}{\partial J_1} 
\\
&= \nu_x  +\frac{J_1}{2\kappa}
-\frac{J_1}{8\kappa^3\nu_x^2}\,\biggl[ (1+n)(1-n)(2-n) -\beta_0^2\,(1+30n-7n^2) \biggr] 
\\
& \quad
-\frac{J_1}{4\kappa^5\nu_x^4} \,
\biggl\{ 6\kappa^4\nu_x^4
+2 \kappa^2\nu_x^2 (1-n)(2-n-3\beta_0^2) 
+\frac{5}{6}(1-n)^2(2-n-3\beta_0^2)^2
\,\biggr\} \,.
\end{split}
\eq
Our real interest is in the value of $\langle x\rangle$. Now 
\bq
\begin{split}
\xi &= \Bigl(\frac{2J}{\kappa\nu_x}\Bigr)^{1/2}\,\cos\phi
\\
&= \Bigl(\frac{2}{\kappa\nu_x}\Bigr)^{1/2}J_1^{1/2}\Bigl(1+\frac{\Delta J_1}{J_1}\Bigr)^{1/2}\,\cos(\phi_1+\Delta\phi_1)
\\
&\simeq \Bigl(\frac{2}{\kappa\nu_x}\Bigr)^{1/2}\bigl(J_1^{1/2}+\frac12J_1^{-1/2}\Delta J_1\bigr)\,
\Bigl[\,\cos\phi_1\cos(\Delta\phi_1) - \sin\phi_1\sin(\Delta\phi_1)\,\Bigr]
\\
&\simeq \Bigl(\frac{2}{\kappa\nu_x}\Bigr)^{1/2}\bigl(J_1^{1/2}+\frac12J_1^{-1/2}\Delta J_1\bigr)\,
\Bigl[\,\cos\phi_1 - \Delta\phi_1\sin\phi_1\,\Bigr]
\\
&\simeq \Bigl(\frac{2J_1}{\kappa\nu_x}\Bigr)^{1/2} \, \cos\phi_1
\\
&\qquad
-\frac{2J_1}{\kappa^2\nu_x^2}\, \biggl[\, \frac{\kappa\nu_x}{4}\,\bigl[\, \cos\phi_1 - \cos(3\phi_1) \,\bigr]
+\frac{1-n}{8\kappa\nu_x}\Bigl(\frac{2-n}{3}-\beta_0^2\Bigr)\,
\bigl[\,3\cos\phi_1 +\cos(3\phi_1) \,\bigr] \biggr]\, \cos\phi_1
\\
&\qquad
-\frac{6J_1}{\kappa^2\nu_x^2}\,\biggl[\,\frac{\kappa\nu_x}{4}\,\bigl[\, \sin\phi_1 - \frac13\,\sin(3\phi_1) \,\bigr]
+\frac{1-n}{8\kappa\nu_x}\Bigl(\frac{2-n}{3}-\beta_0^2\Bigr)\,
\bigl[\,3\sin\phi_1 +\frac13\,\sin(3\phi_1) \,\bigr] \biggr]\, \sin\phi_1 \,.
\end{split}
\eq
We want the average $\langle \xi\rangle$, which is given by the secular terms 
\bq
\begin{split}
\langle \xi\rangle &\simeq -\frac{J_1}{\kappa^3\nu_x^3}\, 
\Bigl[\, \kappa^2\nu_x^2 +\frac{1-n}{2}(2-n-3\beta_0^2) \,\Bigr] 
\\
&= -\frac{J_1}{\kappa^3\nu_x^3}\, 
\Bigl[\, 2-n-\beta_0^2 +\frac{1-n}{2}(2-n-3\beta_0^2) \,\Bigr] \,.
\end{split}
\eq
To this level of approximation
\bq
J_1 \simeq \frac{\kappa\nu_x\,}{2}\,\Bigl(\frac{x_0^2}{r_0^2} + \frac{1}{\kappa^2\nu_x^2}\,\frac{p_{x0}^2}{p_0^2}\Bigr) \,.
\eq
Then to $O(J_1)$, noting that $\kappa\nu_x = \sqrt{2-n-\beta_0^2}$,
\bq
\langle \xi\rangle = -\frac{1}{2(2-n-\beta_0^2)}\, 
\Bigl[\, 2-n-\beta_0^2 +\frac{1-n}{2}(2-n-3\beta_0^2) \,\Bigr] 
\Bigl(\frac{x_0^2}{r_0^2} + \frac{1}{2-n-\beta_0^2}\,\frac{p_{x0}^2}{p_0^2}\Bigr) \,.
\eq

Let us also calculate $\langle p_\xi\rangle$ as a sanity check. 
Because the Hamiltonian is invariant under a change of sign $p_\xi \to -p_\xi$,
we must have $\langle p_\xi\rangle = 0$.
We obtain
\bq
\begin{split}
p_\xi &= -\sqrt{2J\kappa\nu_x}\,\sin\phi
\\
&= -\sqrt{2\kappa\nu_x}\,J_1^{1/2}\Bigl(1+\frac{\Delta J_1}{J_1}\Bigr)^{1/2}\,\sin(\phi_1+\Delta\phi_1)
\\
&\simeq -\sqrt{2\kappa\nu_x}\,\bigl(J_1^{1/2}+\frac12J_1^{-1/2}\Delta J_1\bigr)\,
\Bigl[\,\sin\phi_1\cos(\Delta\phi_1) + \cos\phi_1\sin(\Delta\phi_1)\,\Bigr]
\\
&\simeq -\sqrt{2\kappa\nu_x}\,\bigl(J_1^{1/2}+\frac12J_1^{-1/2}\Delta J_1\bigr)\,\Bigl[\,\sin\phi_1 + \Delta\phi_1\cos\phi_1\,\Bigr]
\\
&\simeq -\sqrt{2J_1\kappa\nu_x} \, \sin\phi_1
\\
&\qquad
+\frac{2J_1}{\kappa\nu_x}\, \biggl[\, \frac{\kappa\nu_x}{4}\,\bigl[\, \cos\phi_1 - \cos(3\phi_1) \,\bigr]
+\frac{1-n}{8\kappa\nu_x}\Bigl(\frac{2-n}{3}-\beta_0^2\Bigr) 
\,\bigl[\,3\cos\phi_1 +\cos(3\phi_1) \,\bigr] \biggr]\, \sin\phi_1
\\
&\qquad
-\frac{6J_1}{\kappa\nu_x}\,\biggl[\,\frac{\kappa\nu_x}{4}\,\bigl[\, \sin\phi_1 - \frac13\,\sin(3\phi_1) \,\bigr]
+\frac{1-n}{8\kappa\nu_x}\Bigl(\frac{2-n}{3}-\beta_0^2\Bigr) 
\,\bigl[\,3\sin\phi_1 +\frac13\,\sin(3\phi_1) \,\bigr] \biggr]\, \cos\phi_1 \,.
\end{split}
\eq
There are no secular terms so $\langle p_\xi\rangle=0$ as required.

\subsection{Helicity at magic momentum}
Using eq.~\eqref{eq:dadt_taylor}, the average over an orbit is
\bq
\label{eq:dadtavg}
\biggl\langle \frac{d\alpha_\beta}{d\theta} \biggr\rangle \simeq 
-\frac{1}{\gamma_0} \bigl[\, 2\langle\xi\rangle +(3-3n-\beta_0^2)\langle\xi^2\rangle \,\bigr] \,.
\eq
We now know that to to  the relevant order
\bq
\xi \simeq \sqrt{\frac{2J_1}{\kappa\nu_x}}\,\cos(\nu_x\phi) 
-\frac{2(2-n-\beta_0^2) +(1-n)(2-n-3\beta_0^2)}{2(2-n-\beta_0^2)}\, 
\,\frac{J_1}{\kappa\nu_x} \,.
\eq
Then the secular term is given by
\bq
\label{eq:dadtsec}
\begin{split}
\biggl\langle \frac{d\alpha_\beta}{d\theta} \biggr\rangle &\simeq 
-\frac{1}{\gamma_0}\,\biggl[\,
-\frac{2(2-n-\beta_0^2) +(1-n)(2-n-3\beta_0^2)}{2-n-\beta_0^2}
+3-3n-\beta_0^2 
\,\biggr]\,\frac{J_1}{\kappa\nu_x} \,.
\end{split}
\eq
Simplify to obtain
\bq
\begin{split}
&\quad -2(2-n-\beta_0^2) -(1-n)(2-n-3\beta_0^2)
+(3-3n-\beta_0^2)(2-n-\beta_0^2)
\\
&= -2(2-n) +2\beta_0^2 -(1-n)(2-n) +3(1-n)\beta_0^2
+3(1-n)(2-n)
-(5-4n)\beta_0^2
+\beta_0^4
\\
&= \beta_0^4 + n\beta_0^2 -4+2n +4-6n+2n^2
\\
&= \beta_0^4 - n(4 -2n -\beta_0^2) \,.
\end{split}
\eq
Hence
\bq
\label{eq:mydadt}
\biggl\langle \frac{d\alpha_\beta}{d\theta} \biggr\rangle \simeq 
-\frac{1}{2\gamma_0}\, \frac{\beta_0^4 - n(4 -2n -\beta_0^2)}{2-n-\beta_0^2} 
\,\Bigl(\frac{x_0^2}{r_0^2} + \frac{1}{2-n-\beta_0^2}\,\frac{p_{x0}^2}{p_0^2}\Bigr) \,.
\eq
Note that in the above calculations, I have averaged only over the angle $\phi$ but not the amplitude.
We can also average over the amplitude, but this is trivial and not necessary at present.
In tracking simulations to be reported below, I employed the initial conditions $x=x_0$ and $p_{x0}=0$.

Some important special cases are $n=0$ (logarithmic potential) and $n=1$ (relativistic Kepler problem).
For brevity I set $p_{x0}=0$. Then
\bq
\label{eq:dadt_n01}
\biggl\langle \frac{d\alpha_\beta}{d\theta} \biggr\rangle \simeq 
\begin{cases}
\displaystyle
-\frac12\,\frac{\gamma_0\beta_0^4}{\gamma_0^2+1}\,\frac{x_0^2}{r_0^2} & \qquad (n=0) \,,
\\
\\
\displaystyle
\frac12\,\frac{2+\beta_0^2}{\gamma_0}\,\frac{x_0^2}{r_0^2} & \qquad (n=1) \,.
\end{cases}
\eq
Let us analyze some limiting cases.
Begin with $a\to\infty$, i.e.~$\beta\to0$.
Then for fixed $n$,
\bq
\lim_{a\to\infty} \biggl\langle \frac{d\alpha_\beta}{d\theta} \biggr\rangle = 
\begin{cases}
\displaystyle
0 & \qquad (n=0) \,,
\\
\displaystyle
\frac{x_0^2}{r_0^2} & \qquad (n\ne 0) \,.
\end{cases}
\eq
Next consider the opposite limit $a\to0$, i.e.~$\beta\to1$ and $\gamma\to\infty$.
Then for all fixed $n$,
\bq
\lim_{a\to0} \biggl\langle \frac{d\alpha_\beta}{d\theta} \biggr\rangle = 0 \,.
\eq

\section{\label{sec:kep} Kepler problem}
I have calculated the spin precession exactly for the relativistic Kepler problem \cite{ManeRelKepler}.
Let us use the exact solution to validate the perturbative solution for the spin decoherence rate on a betatron orbit.
The Hamiltonian for the orbital motion in the relativistic Kepler problem is, in polar coordinates $(r,\theta,z)$
\bq
\label{eq:hkepler}
H_{\rm Kepler} = \sqrt{m^2+p_r^2+\frac{L^2}{r^2}} - \frac{K^2}{r} \,.
\eq
Here $L$ is the orbital angular momentum, and for our application
$K^2 = eE_0r_0^2 = m\gamma_0\beta_0^2r_0$.
Let the energy be $E$.
Using the centripetal acceleration for motion in a circle of radius $r$,
\bq
\frac{\gamma mv^2}{r} = \frac{K^2}{r^2} \,.
\eq
Hence $K^2/r = m\gamma v^2 = m(\gamma - 1/\gamma)$. Then from eq.~\eqref{eq:hkepler}
\bq
E = m \gamma - m\Bigl(\gamma - \frac{1}{\gamma}\Bigr) = \frac{m}{\gamma} \,.
\eq
We treat only on-energy betatron orbits hence $E = m/\gamma_0$.
The secular rate of the spin phase advance, relative to the orbit, is \cite{ManeRelKepler}
\bq
\label{eq:dadtseckep}
\frac{d\alpha}{d\theta} = \frac{aEK^4}{m(L^2-K^4)} -\frac{\sqrt{L^2-K^4}}{L}  \,.
\eq
Consider an orbit with initial conditions $r=r_0+x_0$ and $p_r=0$. 
Using eq.~\eqref{eq:hkepler} with $E=m/\gamma_0$, $p_r=0$ and $r=r_0+x_0$, the angular momentum is given by
\bq
\label{eq:lkepler}
\begin{split}
L^2 &= r^2 \,\biggl[\,\biggl(E + \frac{K^2}{r}\biggr)^2 - m^2\,\biggr]
\\
&= (r_0+x_0)^2 \,\biggl[\,\biggl(\frac{m}{\gamma_0} + \frac{m\gamma_0\beta_0^2r_0}{r_0+x_0}\biggr)^2 - m^2\,\biggr]
\\
&= m^2(r_0+x_0)^2\,
\biggl[\,-\beta_0^2 
+\frac{2\beta_0^2}{1+x_0/r_0}
+\frac{\gamma_0^2\beta_0^4}{(1+x_0/r_0)^2}
\,\biggr]
\\
&= m^2\beta_0^2r_0^2\, \biggl[\,1 -\frac{x_0^2}{r_0^2} +\gamma_0^2\beta_0^2 \,\biggr]
\\
&= m^2\beta_0^2\gamma_0^2r_0^2\, \biggl( 1 -\frac{1}{\gamma_0^2}\,\frac{x_0^2}{r_0^2} \biggr) \,.
\\
\end{split}
\eq
Then
\bq
\begin{split}
L^2 -K^4 &= m^2\beta_0^2\gamma_0^2r_0^2\, \biggl[\,1 -\beta_0^2 -\frac{1}{\gamma_0^2}\,\frac{x_0^2}{r_0^2} \,\biggr] 
\\
&= m^2\beta_0^2r_0^2\, \biggl( 1 - \frac{x_0^2}{r_0^2} \biggr) \,.
\end{split}
\eq
Substituting in eq.~\eqref{eq:dadtseckep}, the spin decoherence rate is, 
using the magic gamma condition $a=1/(\beta_0^2\gamma_0^2)$
\bq
\begin{split}
\frac{d\alpha}{d\theta} &= \frac{m^2\beta_0^2r_0^2/\gamma_0}{m^2\beta_0^2r_0^2} \biggl( 1 - \frac{x_0^2}{r_0^2} \biggr)^{-1}
- \frac{m\beta_0r_0}{m\beta_0\gamma_0r_0}\, \biggl( 1 - \frac{x_0^2}{r_0^2} \biggr)^{1/2}
\biggl( 1 -\frac{1}{\gamma_0^2}\,\frac{x_0^2}{r_0^2} \biggr)^{-1/2}
\\
&\simeq \frac{1}{\gamma_0}\,\biggl[\, 1 + \frac{x_0^2}{r_0^2}
- \biggl( 1 -\frac{\gamma_0^2-1}{2\gamma_0^2}\,\frac{x_0^2}{r_0^2} \biggr)
\,\biggr]
\\
&= \frac12\,\frac{2+\beta_0^2}{\gamma_0}\,\frac{x_0^2}{r_0^2} \,.
\end{split}
\eq
This confirms the result in eq.~\eqref{eq:dadt_n01}, which was derived using perturbation theory.

\section{\label{sec:syncosc} Synchrotron oscillations}
In Section \ref{sec:offenergy} I calculated the spin decoherence rate for motion on off-energy orbits
(see eq.~\eqref{eq:dadthetaoffenergyavg}).
In that section, there was no rf cavity and the total energy was a dynamical invariant.
Here I extend the calculation to include an rf cavity and synchrotron oscillations.
The spin decoherence rate is still given by eq.~\eqref{eq:dadthetaoffenergyavg},
but now $\lambda_p$ oscillates and the statictical averages in that formula must be derived.
To do so I shall perform a canonical transformation to separate the radial motion into betatron and dispersion terms.
This will require the determination of the second order dispersion.
I shall then diagonalize the Hamiltonian using action-angle variables to determine the relevant averages for the synchrotron oscillations.
The results are in good agreement with those from tracking simulations.
This section is self-contained.
I treat a ring of arbitrary structure, all-magnetic or all-electric, but without transverse coupling.

\subsection{Basic Hamiltonian}
I treat a particle of mass $m$ and charge $e$ and I set $c=1$ below.
For the formal treatment, we can treat a ring with both electric and magnetic fields.
The indepenent variable is the arc-length $s$ along the reference orbit.
The ring circumference is $2\pi R$ and I define the generalized azimuth $\theta = s/R$.
The curvature of the reference orbit is $1/\rho_0$ and is zero in the straight sections.
The dynamical variables are $(x,p_x,z,p_z,-t,H)$.
I assume there is only one (zero length) rf cavity, localized at $\theta=\theta_{\rm rf}$.
I treat only a stationary rf bucket, and I assume $eV_0>0$.
The Hamiltonian is
\bq
\begin{split}
K &= -\biggl(1+\frac{x}{\rho_0}\biggr)\biggl[\,(H-\Phi)^2 - m^2 - p_x^2 - p_z^2\,\biggr]^{1/2} 
-eA_s 
\\
&\quad
+\frac{eV_0}{R\omega_{\rm rf}}\,\cos(\omega_{\rm rf}(t-t_*)) \,\delta_p(\theta-\theta_{\rm rf}) \,.
\end{split}
\eq
Here $\omega_{\rm rf} = h\omega_0 = h\beta_0/R$,
$t_* = s/\beta_0$ is the time of arrival of the reference particle
and $\delta_p$ is the periodic $\delta$ function
\bq
\delta_p(\theta-\theta_{\rm rf}) = \sum_{j=-\infty}^\infty \delta(\theta-\theta_{\rm rf}-2j\pi) \,.
\eq
We perform a canonical transformation to subtract the time of flight of the reference particle
to obtain $\tau = s/\beta_0 -t = t_*-t$.
We also subtract the reference energy $H_0$ to obtain an energy offset $\Delta H = H - H_0$.
The generating function is (ignoring variables for which the transformation is the identity)
\bq
G_\Delta = \frac{s\Delta H}{\beta_0} -(H_0+\Delta H)t \,.
\eq
Then
\begin{subequations}
\begin{align}
\tau &= \frac{\partial G_\Delta}{\partial (\Delta H)} = \frac{s}{\beta_0} - t \,,
\\
H &= -\frac{\partial G_\Delta}{\partial t} = H_0+\Delta H \,.
\end{align}
\end{subequations}
The transformed Hamiltonian is, using $p_0 = H_0\beta_0$ 
\bq
\begin{split}
K_1 &= K + \frac{\partial G_\Delta}{\partial s}
\\
&= \frac{\Delta H}{\beta_0} 
-\biggl(1+\frac{x}{\rho_0}\biggr)\biggl[\,(H_0+\Delta H-\Phi)^2 - m^2 - p_x^2 - p_z^2\,\biggr]^{1/2} 
\\
&\quad
-eA_s 
+\frac{eV_0}{R\omega_{\rm rf}}\,\cos(\omega_{\rm rf}\tau) \,\delta_p(\theta-\theta_{\rm rf}) \,.
\end{split}
\eq
We divide all the momenta by the reference momentum $p_0$ and divide the Hamiltonian by $p_0$ also.
This preserves the Hamiltonian structure of the equations
\bq
\frac{dq}{ds} = \frac{\partial (K_1/p_0)}{\partial (p/p_0)} \,,\qquad
\frac{d(p/p_0)}{ds} = -\frac{\partial (K_1/p_0)}{\partial q} \,.
\eq
Let $\bar{p}_x=p_x/p_0$ and $\bar{p}_z=p_z/p_0$.
The transformed Hamiltonian is
\bq
\begin{split}
K_2 = \frac{K_1}{p_0} &= \frac{\Delta H}{p_0\beta_0} 
-\biggl(1+\frac{x}{\rho_0}\biggr)
\biggl[\,\frac{H_0^2}{p_0^2}\Bigl(1+\frac{\Delta H}{H_0}-\frac{\Phi}{H_0}\Bigr)^2 - \frac{m^2}{p_0^2} - \bar{p}_x^2 - \bar{p}_z^2\,\biggr]^{1/2} 
\\
&\quad
-\frac{eA_s}{p_0} 
+\frac{eV_0}{p_0R\omega_{\rm rf}}\,\cos(\omega_{\rm rf}\tau) \,\delta_p(\theta-\theta_{\rm rf}) \,.
\end{split}
\eq
Next we transform $(\Delta H)/p_0$ to $\lambda_p = (\Delta H)/(p_0\beta_0) = (1/\beta_0^2)(\Delta H/H_0)$.
This is a scaling transformation and also requires a scaling of the conjugate variable to $\sigma = \beta_0\tau$.
The generating function is (ignoring variables for which the transformation is the identity)
\bq
G_{\beta_0} = \beta_0 \tau \lambda_p \,.
\eq
Then
\begin{subequations}
\begin{align}
\sigma &= \frac{\partial G_{\beta_0}}{\partial \lambda_p} = \beta_0\tau \,,
\\
\frac{\Delta H}{p_0} &= \frac{\partial G_{\beta_0}}{\partial \tau} = \beta_0\lambda_p \,.
\end{align}
\end{subequations}
It is also convenient to define a scaled electrostatic potential
\bq
\Psi = \frac{\Phi}{p_0\beta_0} = \frac{\Phi}{H_0\beta_0^2} \,.
\eq
Then the transformed Hamiltonian is
\bq
\begin{split}
K_3(\sigma,\lambda_p) = K_2\Bigl(\tau,\frac{\Delta H}{p_0}\Bigr)
&= \lambda_p
-\biggl(1+\frac{x}{\rho_0}\biggr)
\biggl[\,\frac{(1+\beta_0^2(\lambda_p-\Psi))^2}{\beta_0^2} - \frac{1}{\beta_0^2\gamma_0^2} - \bar{p}_x^2 - \bar{p}_z^2\,\biggr]^{1/2} 
\\
&\quad
-\frac{eA_s}{p_0} 
+\frac{eV_0}{p_0R\omega_{\rm rf}}\,\cos(\frac{h\sigma}{R}) \,\delta_p(\theta-\theta_{\rm rf}) 
\\
&= \lambda_p
-\biggl(1+\frac{x}{\rho_0}\biggr)
\biggl[\,1+2(\lambda_p-\Psi) +\beta_0^2(\lambda_p-\Psi)^2 - \bar{p}_x^2 - \bar{p}_z^2\,\biggr]^{1/2} 
\\
&\quad
-\frac{eA_s}{p_0} 
+\frac{eV_0}{p_0R\omega_{\rm rf}}\,\cos(\frac{h\sigma}{R}) \,\delta_p(\theta-\theta_{\rm rf}) \,.
\end{split}
\eq
One must now perform a further canonical transformation to express the 
radial motion in betatron and dispersion terms.
We write $x = x_\beta+x_d$ and $p_x=p_{x\beta}+p_{xd}$, with an obvious notation.
Suppose also that $(z,p_z,\sigma,\lambda_p)$ are transformed to
$(\hat{z},\hat{p}_z,\hat{\sigma},\mu)$.
We employ the generating function
(here $\lambda$ is a dummy variable of integration)
\bq
\label{eq:gbeta}
G_\beta = \sigma \mu +z\hat{p}_z 
+(x-x_d)p_{x\beta} + xp_{xd} -\int_0^{\mu} x_d\,\frac{\partial p_{xd}}{\partial \lambda}\,d\lambda \,.
\eq
Note that the lower limit of the integral may not always be zero.
It could be $-\infty$ or any other value.
The lower limit should be chosen so that the contribution to the integral is zero.
Then
\begin{subequations}
\begin{align}
x_\beta &= x-x_d \,,
\\
p_x &= p_{x\beta} + p_{xd} \,,
\\
\hat{z} &= z \,,
\\
p_z &= \hat{p}_z \,,
\\
\lambda_p &= \mu \,,
\\
\hat{\sigma} &= \sigma -p_{x\beta}\,\frac{\partial x_d}{\partial\mu}
+(x-x_d)\,\frac{\partial p_{xd}}{\partial\mu} 
\nonumber\\
&= \sigma -p_{x\beta}\,\frac{\partial x_d}{\partial\mu}
+x_\beta\,\frac{\partial p_{xd}}{\partial\mu} \,.
\end{align}
\end{subequations}
The last line indicates the presence of synchrobetatron coupling,
because $\hat{\sigma}$ depends on the betatron variables $x_\beta$ and $p_{x\beta}$.
The transformed Hamiltonian is 
\bq
\begin{split}
K_{sb} &= K_3 + \frac{\partial G_\beta}{\partial s}
\\
&= K_3
-p_{x\beta}\,\frac{\partial x_d}{\partial s} 
+(x_\beta+x_d)\,\frac{\partial p_{xd}}{\partial s} 
-\frac{\partial\ }{\partial s}\biggl(
\int_0^{\mu} x_d\,\frac{\partial p_{xd}}{\partial \lambda}\,d\lambda \biggr) \,.
\end{split}
\eq
To proceed further we need to expand the square root in $K_3$ 
and all the terms in the potentials, up to the desired order.
It is, however, possible to diagonalize the above Hamiltonian for the synchrotron oscillations
(using action-angle variables) via a formal procedure.

\subsection{Action-angle variables for synchrotron oscillations}
To derive action-angle variables for the synchrotron oscillations,
it is possible to treat both all-magnetic and all-electric 
in a unified formal procedure as follows. 
We set the betatron terms to zero, so $\hat{\sigma} = \sigma$
and treat only the dispersion terms below.
It is convenient to smooth out the rf cavity around the ring,
i.e.~to replace the periodic $\delta$-function by $1/(2\pi)$.
Expanding the cosine, the Hamiltonian yields, for the synchrotron motion,
\bq
K_4 \simeq \textrm{const}
-\frac{eV_0h}{2\pi p_0\beta_0R^2}\,\frac{\sigma^2}{2} 
+\frac{eV_0h^3}{2\pi p_0\beta_0R^4}\, \frac{\sigma^4}{24} 
-\frac{\bar{a}}{2}\,\lambda_p^2 -\frac{\bar{b}}{6}\,\lambda_p^3
+\cdots
\eq
Here $\bar{a}$ and $\bar{b}$ are formal constants,
i.e.~they do not depend on the dynamical variables
and their explicit expressions in terms of the dispersions are not required below.
It is also convenient to change the independent variable to $\theta$.
Also, it is preferable for the quadratic terms in the Hamiltonian to be positive definite.
Hence we scale $\bar{\sigma} = -\sigma/R$ and reverse the sign of the Hamiltonian.
This preserves the Hamiltonian structure of the equations
\begin{subequations}
\begin{align}
\frac{d\bar{\sigma}}{d\theta} &= -\frac{d\sigma}{ds} = \frac{\partial (-K_4)}{\partial \lambda_p} \,,
\\
\frac{d\lambda_p}{d\theta} &= R\,\frac{d\lambda_p}{ds} 
= -R\,\frac{\partial K_4}{\partial \sigma} 
= -\frac{\partial (-K_4)}{\partial \bar{\sigma}} \,.
\end{align}
\end{subequations}
Then, dropping the constant term,
\bq
K_5 = \frac{\bar{a}}{2}\,\lambda_p^2 +\frac{\bar{b}}{6}\,\lambda_p^3 
+\frac12\,\frac{eV_0h}{2\pi\beta_0p_0}\,\bar{\sigma}^2
-\frac{1}{24}\frac{eV_0h^3}{2\pi\beta_0p_0}\,\bar{\sigma}^4 \,.
\eq
We can diagonalize this formally.
The small amplitude synchrotron tune is given by
\bq
\nu_s^2 = \frac{eV_0h}{2\pi\beta_0p_0}\,\bar{a} \,.
\eq
We define action-angle variables via
\bq
\lambda_p = \sqrt{\frac{2J_1\nu_s}{\bar{a}}}\,\cos\phi_1 \,,\qquad
\bar{\sigma} = \sqrt{\frac{2J_1\nu_s}{eV_0h/(2\pi\beta_0p_0)}}\,\sin\phi_1 \,.
\eq
Then
\bq
\begin{split}
K_5 &= \nu_s\,J_1\cos^2\phi_1  
+\frac{\bar{b}}{6}\,(2J_1\nu_s/\bar{a})^{3/2}\,\cos^3\phi_1 
+\nu_s\,J_1\sin^2\phi_1
+O(J_1^2) 
\\
&= \nu_s\,J_1  
+\frac{\bar{b}}{24\bar{a}^{3/2}}\,(2J_1\nu_s)^{3/2}\,\Bigl[\cos(3\phi_1) + 3\cos\phi_1\,\Bigr]
+O(J_1^2) \,.
\end{split}
\eq
We seek to diagonalize this,
hence we perform a canonical transformation to variables $(J_2,\phi_2)$ to eliminate the oscillating terms.
A suitable generating function is
\bq
G = \phi_1 J_2 
-\frac{\bar{b}\nu_s^{1/2}}{24\bar{a}^{3/2}}\,(2J_2)^{3/2}\,\Bigl[\frac13\sin(3\phi_1) + 3\sin\phi_1\,\Bigr] \,.
\eq
Then
\begin{subequations}
\begin{align}
J_1 = \frac{\partial G}{\partial\phi_1} &=
J_2 -\frac{\bar{b}\nu_s^{1/2}}{24\bar{a}^{3/2}}\,(2J_2)^{3/2}\,\Bigl[\,\cos(3\phi_1) + 3\cos\phi_1\,\Bigr] \,,
\\
\phi_2 = \frac{\partial G}{\partial J_2} &=
\phi_1  
-\frac{\bar{b}\nu_s^{1/2}}{8\bar{a}^{3/2}}\,(2J_2)^{1/2}\,\Bigl[\frac13\sin(3\phi_1) + 3\sin\phi_1\,\Bigr] \,.
\end{align}
\end{subequations}
Then $K_5 = \nu_sJ_2 + O(J_2^2)$.
Now
\bq
\begin{split}
\lambda_p &\propto J_1^{1/2}\cos\phi_1 
\\
&\simeq 
\biggl[\,
J_2 -\frac{\bar{b}\nu_s^{1/2}}{24\bar{a}^{3/2}}\,(2J_2)^{3/2}\,\Bigl(\cos(3\phi_2) + 3\cos\phi_2\Bigr) 
\biggr]^{1/2}\,
\cos\Bigl(
\phi_2
+\frac{\bar{b}\nu_s^{1/2}}{8\bar{a}^{3/2}}\,(2J_2)^{1/2}\,\Bigl[\frac13\sin(3\phi_2) + 3\sin\phi_2\,\Bigr]
\Bigr)
\\
&\simeq J_2^{1/2}
\biggl[\,1 
-\frac{\bar{b}\nu_s^{1/2}}{24\bar{a}^{3/2}}\,(2J_2)^{1/2}\,\Bigl(\cos(3\phi_2) + 3\cos\phi_2\Bigr) 
\biggr]\, 
\biggl\{
\cos\phi_2
-\sin\phi_2\,
\frac{\bar{b}\nu_s^{1/2}}{8\bar{a}^{3/2}}\,(2J_2)^{1/2}\,\Bigl[\frac13\sin(3\phi_2) + 3\sin\phi_2\,\Bigr]
\biggr\}
\\
&\simeq J_2^{1/2}\cos\phi_2 -\frac{1}{2\sqrt2}\,\frac{\bar{b}\nu_s^{1/2}}{\bar{a}^{3/2}}\,J_2 +\cdots
\end{split}
\eq
Then
\bq
\langle \lambda_p \rangle \simeq 
-\frac{\bar{b}\nu_s^{1/2}}{2\sqrt2\bar{a}^{3/2}}\,\sqrt{\frac{2\nu_s}{\bar{a}}}\,   \langle J_2\rangle 
\simeq
-\frac{\bar{b}}{4\bar{a}^2}\,
\biggl(\bar{a}\,\lambda_{p0}^2 +\frac{eV_0h}{2\pi\beta_0p_0}\,\bar{\sigma}_0^2 \biggr)\,.
\eq
Also
\bq
\langle \lambda_p^2 \rangle \simeq \frac{\nu_s}{\bar{a}}\, \langle J_2\rangle 
\simeq
\frac{1}{2\bar{a}}\,
\biggl(\bar{a}\,\lambda_{p0}^2 +\frac{eV_0h}{2\pi\beta_0p_0}\,\bar{\sigma}_0^2 \biggr)\,.
\eq
We need to determine $\bar{a}$ and $\bar{b}$ explicitly.
Then the expressions for $\langle \lambda_p \rangle$ and $\langle \lambda_p^2 \rangle$
can be employed in eq.~\eqref{eq:dadthetaoffenergyavg}.
Note that if $\bar{\sigma}_0=0$ then the value of $eV_0h$ is not required.

\subsection{Time of flight}
Before performing the canonical transformation to separate the radial motion into betatron and dispersion terms,
let us first derive an expression for the time of flight, essentially the frequency slip factor.
Only motion in the horizontal plane will be treated.
The dispersion functions are defined via the fixed point of the one-turn map, for the radial motion of an off-energy orbit.
We write
$x_d = D_1 \lambda_p + D_2 \lambda_p^2 + \cdots$.
There is also a variable $p_{xd}$, which will be determined below.
The scaled potential is (in the median plane)
\bq
\Psi = \frac{\Phi}{H_0\beta_0^2} = \Psi_{,x}x +\frac{\Psi_{,xx}}{2}\,x^2 +\frac{\Psi_{,xxx}}{6}\,x^3 +\cdots
\eq
Note that $\Psi=0$ in an all-magnetic ring.
Also $\Psi_{,x} = 1/\rho_0$ in an all-electric ring.
The time of flight for a small arc-length $\delta s$ is $\delta t = \delta L/v$.
Now 
\bq
\frac{1}{\beta} = \frac{H-\Phi}{\sqrt{(H-\Phi)^2-m^2}} \,.
\eq
Hence
\bq
\begin{split}
\frac{\beta_0}{\beta} &= \frac{\beta_0 H_0(1+\beta_0^2(\lambda_p-\Psi))}{\sqrt{H_0^2(1+\beta_0^2(\lambda_p-\Psi))^2-m^2}}
\\
&= \frac{1+\beta_0^2(\lambda_p-\Psi)}{\sqrt{1+2\lambda_p-\Psi+\beta_0^2(\lambda_p-\Psi)^2}}
\\
&= \frac{1+\beta_0^2(\lambda_p-\Psi)}{\sqrt{(1+\lambda_p-\Psi)^2-(\lambda_p-\Psi)^2/\gamma_0^2}}
\\
&\simeq \frac{1+\beta_0^2(\lambda_p-\Psi)}{(1+\lambda_p-\Psi)\sqrt{1-(\lambda_p-\Psi)^2/\gamma_0^2}}
\\
&\simeq \Bigl[\,1+\beta_0^2(\lambda_p-\Psi)\,\Bigr] 
\Bigl[\,1-\lambda_p+\Psi +(\lambda_p-\Psi)^2\,\Bigr] 
\,\biggl[\, 1 +\frac{(\lambda_p-\Psi)^2}{2\gamma_0^2}\,\biggr]
\\
&\simeq 1 -\frac{\lambda_p-\Psi}{\gamma_0^2} +\frac{3(\lambda_p-\Psi)^2}{2\gamma_0^2} \,.
\end{split}
\eq
Then 
\bq
\begin{split}
\beta_0\,\frac{dt}{ds} &= \frac{\beta_0}{\beta}\sqrt{\frac{r^2}{r_0^2} + \Bigl(\frac{dr}{ds}\Bigr)^2}
\\
&\simeq \sqrt{\Bigl(1+\frac{D_1}{\rho_0}\,\lambda_p+\frac{D_2}{\rho_0}\,\lambda_p^2\Bigr)^2 + D_1^{\prime2}\lambda_p^2} \;
\biggl(1 -\frac{\lambda_p-\Psi}{\gamma_0^2} +\frac{3(\lambda_p-\Psi)^2}{2\gamma_0^2} \biggr)
\\
&\simeq \biggl(1+\frac{D_1}{\rho_0}\,\lambda_p+\frac{D_2}{\rho_0}\,\lambda_p^2 +\frac{D_1^{\prime2}}{2}\lambda_p^2\biggr)
\biggl(1 -\frac{\lambda_p-\Psi}{\gamma_0^2} +\frac{3(\lambda_p-\Psi)^2}{2\gamma_0^2} \biggr)
\\
&\simeq 1+\frac{D_1}{\rho_0}\,\lambda_p -\frac{\lambda_p-\Psi}{\gamma_0^2}
+\frac{D_2}{\rho_0}\,\lambda_p^2 
+\frac{D_1^{\prime2}}{2}\lambda_p^2
+\frac{3(\lambda_p-\Psi)^2}{2\gamma_0^2} 
-\frac{D_1(\lambda_p-\Psi)}{\gamma_0^2\rho_0} \,.
\end{split}
\eq
Recall that $\Psi=0$ in an all-magnetic ring, so in that case
\bq
\label{eq:tofallmag}
\begin{split}
\frac{d\sigma}{ds} &= 1 - \beta_0\,\frac{dt}{ds} 
\\
&\simeq 
\Bigl(\frac{1}{\gamma_0^2}-\frac{D_1}{\rho_0}\Bigr)\,\lambda_p 
-\biggl[\,\frac{D_2}{\rho_0} 
-\frac{D_1}{\gamma_0^2\rho_0} 
+\frac{D_1^{\prime2}}{2}
+\frac{3}{2\gamma_0^2}\biggr]\,\lambda_p^2 \,.
\end{split}
\eq
Next, $\Psi_{,x}=1/\rho_0$ in an all-electric ring, so in that case
\bq
\label{eq:tofallelec}
\begin{split}
\frac{d\sigma}{ds} 
&\simeq \biggl[\,\frac{1}{\gamma_0^2} -(2-\beta_0^2)\frac{D_1}{\rho_0}\,\lambda_p\,\biggr]\,\lambda_p
\\
&\quad
-\biggl[ (2-\beta_0^2)\frac{D_2}{\rho_0}
+\frac{\Psi_{,xx}D_1^2}{2\gamma_0^2}
+\frac{D_1^{\prime2}}{2}
+\frac{3(1-D_1/\rho_0)^2}{2\gamma_0^2} 
-\frac{D_1(1-D_1/\rho_0)}{\gamma_0^2\rho_0} \,\biggr]\, \lambda_p^2 
\\
&\simeq \biggl[\,\frac{1}{\gamma_0^2} -(2-\beta_0^2)\frac{D_1}{\rho_0}\,\lambda_p \,\biggr]\,\lambda_p
\\
&\quad
-\biggl[ (2-\beta_0^2)\frac{D_2}{\rho_0}
+\frac{D_1^{\prime2}}{2}
-\frac{4D_1}{\gamma_0^2\rho_0} 
+\frac{5D_1^2}{2\gamma_0^2\rho_0^2} 
+\frac{\Psi_{,xx}D_1^2}{2\gamma_0^2}
+\frac{3}{2\gamma_0^2} 
\,\biggr]\, \lambda_p^2 \,.
\end{split}
\eq
These expressions will be used below, to cross-check the results from the canonical transformations.

\subsection{Betatron and dispersion motion: all-magnetic ring}
An rf cavity is not required, 
to express the Hamiltonian in betatron and dispersion terms,
hence I shall drop the rf cavity below.
For clarity of the notation, I shall drop the bar on $\bar{p}_x$ and write simply $p_x$ below.
Subscripts `$\beta$' and `$d$' will denote betatron and dispersion terms, respectively.
Only motion in the median plane will be treated.
The Hamiltonian for an all-magnetic ring is, up to third order terms,
\bq
\begin{split}
K_m &= \lambda_p
-\Bigl(1+\frac{x}{\rho_0}\Bigr)\sqrt{1 +2\lambda_p+\beta_0^2\lambda_p^2 -p_x^2} 
+\frac{x}{\rho_0} +\frac{K_x}{2}\,x^2 +\frac{S_x}{6}\,x^3
\\
&= \lambda_p
-\Bigl(1+\frac{x}{\rho_0}\Bigr)\sqrt{(1 +\lambda_p)^2-\lambda_p^2/\gamma_0^2 -p_x^2} 
+\frac{x}{\rho_0} +\frac{K_x}{2}\,x^2 +\frac{S_x}{6}\,x^3
\\
&\simeq \lambda_p
-\Bigl(1+\frac{x}{\rho_0}\Bigr)\,\biggl[\, 1 +\lambda_p
-\frac12\frac{\lambda_p^2/\gamma_0^2 +p_x^2}{1 +\lambda_p} \,\biggr]
+\frac{x}{\rho_0} +\frac{K_x}{2}\,x^2 +\frac{S_x}{6}\,x^3
\\
&\simeq \lambda_p
-\Bigl(1+\frac{x}{\rho_0}\Bigr)\,\biggl[\, 1 +\lambda_p
-\frac{\lambda_p^2}{2\gamma_0^2} 
+\frac{\lambda_p^3}{2\gamma_0^2} 
-(1-\lambda_p)\frac{p_x^2}{2}
\,\biggr]
+\frac{x}{\rho_0} +\frac{K_x}{2}\,x^2 +\frac{S_x}{6}\,x^3
\\
&\simeq -1 +\frac{\lambda_p^2}{2\gamma_0^2} -\frac{\lambda_p^3}{2\gamma_0^2} 
+\Bigl(1-\lambda_p +\frac{x}{\rho_0}\Bigr)\frac{p_x^2}{2}
-\frac{x}{\rho_0}\,\Bigl(\lambda_p -\frac{\lambda_p^2}{2\gamma_0^2} \Bigr)
+\frac{K_x}{2}\,x^2 +\frac{S_x}{6}\,x^3 \,.
\end{split}
\eq
Hamilton's equations for $x$ and $p_x$ are
\begin{subequations}
\begin{align}
\frac{dx}{ds} &= \Bigl(1-\lambda_p+\frac{x}{\rho_0}\Bigr)\,p_x \,,
\\
\frac{dp_x}{ds} &= -\frac{p_x^2}{2\rho_0} 
+\frac{\lambda_p}{\rho_0} 
-\frac{\lambda_p^2}{2\gamma_0^2\rho_0} 
-K_x x -\frac{S_x}{2}\,x^2 \,.
\end{align}
\end{subequations}
Hence the dispersion motion satisfies
\begin{subequations}
\begin{align}
x_d^\prime &= \Bigl(1-\lambda_p+\frac{x_d}{\rho_0}\Bigr)\,p_{xd} \,,
\\
p_{xd}^\prime &= 
-\frac{p_{xd}^2}{2\rho_0} +\frac{\lambda_p}{\rho_0} -\frac{\lambda_p^2}{2\gamma_0^2\rho_0} -K_x x_d -\frac{S_x}{2}\,x_d^2 \,.
\end{align}
\end{subequations}
The second order differential equation is
\bq
\begin{split}
x_d^\pprime &= \frac{x_d^\prime}{\rho_0}\,p_{xd} 
+\Bigl(1 - \lambda_p +\frac{x_d}{\rho_0}\Bigr)\,p_{xd}^\prime
+x_d\,p_{xd} \Bigl(\frac{1}{\rho_0}\Bigr)^\prime
\\
&\simeq
\Bigl(1 - \lambda_p +\frac{x_d}{\rho_0}\Bigr)\,
\biggl[\,
\frac{x_d^{\prime2}}{2\rho_0}
+\frac{\lambda_p}{\rho_0} -\frac{\lambda_p^2}{2\gamma_0^2\rho_0} -K_x x_d -\frac{S_x}{2}\,x_d^2 
\,\biggr]
+x_d x_d^\prime \, \Bigl(\frac{1}{\rho_0}\Bigr)^\prime \,.
\end{split}
\eq
Equating the first order terms yields the well-known equation
\bq
D_1^\pprime + K_xD_1 = \frac{1}{\rho_0} \,.
\eq
Equating the second order terms yields 
\bq
\begin{split}
D_2^\pprime +K_x D_2 &= -\frac{1}{2\gamma_0^2\rho_0}
+\frac{D_1^{\prime2}}{2\rho_0} 
-\frac{S_x}{2}\,D_1^2
+\Bigl(\frac{D_1}{\rho_0}-1\Bigr) D_1^\pprime 
+D_1 D_1^\prime \, \Bigl(\frac{1}{\rho_0}\Bigr)^\prime \,.
\end{split}
\eq
Note also that
\bq
p_{xd}^\prime = \Bigl( x_d^\pprime - \frac{x_d^\prime}{\rho_0}\,p_{xd} -x_d\,p_{xd} \Bigl(\frac{1}{\rho_0}\Bigr)^\prime \Bigr)
\Bigl(1 - \lambda_p +\frac{x_d}{\rho_0}\Bigr)^{-1} \,.
\eq
Hence to the third order
\bq
\begin{split}
x_d p_{xd}^\prime &= x_d\Bigl( x_d^\pprime - \frac{x_d^\prime}{\rho_0}\,p_{xd} -x_d p_{xd} \Bigl(\frac{1}{\rho_0}\Bigr)^\prime \Bigr)
\Bigl(1 - \lambda_p +\frac{x_d}{\rho_0}\Bigr)^{-1}
\\
&\simeq x_d x_d^\pprime \Bigl(1 + \lambda_p -\frac{x_d}{\rho_0}\Bigr) - \frac{x_d x_d^{\prime2}}{\rho_0} 
-x_d^2\,x_d^\prime \Bigl(\frac{1}{\rho_0}\Bigr)^\prime 
\\
&\simeq 
D_1D_1^\pprime \,\lambda_p^2
+\biggl[\, D_2 D_1^\pprime 
+D_1 D_2^\pprime 
-D_1D_1^\pprime \Bigl(\frac{D_1}{\rho_0}-1\Bigr) 
-\frac{D_1D_1^{\prime2}}{\rho_0} \,\biggr] \,\lambda_p^3 
-D_1^2\,D_1^\prime \Bigl(\frac{1}{\rho_0}\Bigr)^\prime \,.
\end{split}
\eq
Also to the third order
\bq
\begin{split}
x_d^\prime p_{xd} &= x_d^{\prime2}\Bigl(1 - \lambda_p +\frac{x_d}{\rho_0}\Bigr)^{-1}
\\
&\simeq x_d^{\prime2}\Bigl(1 + \lambda_p -\frac{x_d}{\rho_0}\Bigr)
\\
&\simeq D_1^{\prime2} \,\lambda_p^2
+\biggl[\, 2D_1^\prime D_2^\prime -D_1^{\prime2}\Bigl(\frac{D_1}{\rho_0} -1\Bigr) \,\biggr] \,\lambda_p^3 \,.
\end{split}
\eq
Next
\bq
\begin{split}
p_{xd} &\simeq x_d^\prime \, \Bigl( 1 +\lambda_p -\frac{x_d}{\rho_0} \Bigr)
\\
&\simeq (D_1^\prime \lambda_p + D_2^\prime \lambda_p^2) \, \Bigl( 1 +\lambda_p -\lambda_p\,\frac{D_1}{\rho_0} \Bigr)
\\
&\simeq D_1^\prime \lambda_p + D_2^\prime \lambda_p^2 + D_1^\prime \lambda_p^2 - \frac{D_1D_1^\prime}{\rho_0} \lambda_p^2 \,.
\end{split}
\eq
Then to the required order
\bq
\begin{split}
x_d\,\frac{\partial p_{xd}}{\partial \lambda_p} &= (D_1 \lambda_p + D_2 \lambda_p^2)\,
\Bigl( D_1^\prime + 2D_2^\prime \lambda_p + 2D_1^\prime \lambda_p - \frac{2D_1D_1^\prime}{\rho_0} \lambda_p \Bigr)
\\
&\simeq  D_1 D_1^\prime \lambda_p + \biggl[\, D_2 D_1^\prime + 2D_1 D_2^\prime +2D_1 D_1^\prime -\frac{2D_1^2D_1^\prime}{\rho_0}\,\biggr]\, \lambda_p^2 \,.
\end{split}
\eq
Hence to the required order (see eq.~\eqref{eq:gbeta})
\bq
\begin{split}
\label{eq:deftildeg}
\tilde{G}_m &= -\int_0^{\lambda_p} x_d\,\frac{\partial p_{xd}}{\partial \lambda}\,d\lambda 
\\
&= -D_1 D_1^\prime \,\frac{\lambda_p^2}{2} 
- \biggl[\, D_2 D_1^\prime + 2D_1 D_2^\prime +2D_1 D_1^\prime -\frac{2D_1^2D_1^\prime}{\rho_0}\,\biggr]\, \frac{\lambda_p^3}{3} \,.
\end{split}
\eq
The transformed Hamiltonian is, dropping the constant
\bq
\begin{split}
K_\beta &\simeq \frac{\partial\tilde{G}_m}{\partial s} -x_d^\prime p_{x\beta} +(x_\beta+x_d)p_{xd}^\prime 
\\
&\quad
+\frac{\lambda_p^2}{2\gamma_0^2} -\frac{\lambda_p^3}{2\gamma_0^2} 
+\Bigl(1-\lambda_p +\frac{x_\beta+x_d}{\rho_0}\Bigr)\frac{(p_{x\beta}+p_{xd})^2}{2}
\\
&\quad
-\frac{x_\beta+x_d}{\rho_0}\,\Bigl(\lambda_p -\frac{\lambda_p^2}{2\gamma_0^2} \Bigr)
+\frac{K_x}{2}\,(x_\beta+x_d)^2 +\frac{S_x}{6}\,(x_\beta+x_d)^3 
\\
&\simeq 
\Bigl(1+\frac{x_\beta}{\rho_0}\Bigr)\frac{p_{x\beta}^2}{2}
+\frac{K_x}{2}\,x_\beta^2 +\frac{S_x}{6}\,x_\beta^3 
\\
&\quad
+\frac{p_{x\beta}^2}{2}\,\Bigl(\frac{x_d}{\rho_0} -\lambda_p\Bigr)
+\frac{x_\beta}{\rho_0}\,p_{x\beta}p_{xd}
+\frac{S_x}{2}\,x_\beta^2x_d
\\
&\quad
+p_{x\beta}\,\biggl[\, -x_d^\prime +\Bigl(1-\lambda_p +\frac{x_d}{\rho_0}\Bigr)p_{xd} \,\biggr]
\\
&\quad
+x_\beta\,\biggl[\, p_{xd}^\prime +\frac{p_{xd}^2}{2\rho_0}
-\frac{\lambda_p}{\rho_0} 
+\frac{\lambda_p^2}{2\gamma_0^2\rho_0}
+K_xx_d 
+\frac{S_x}{2}\,x_d^2 
\,\biggr]
\\
&\quad
+\frac{\partial\tilde{G}_m}{\partial s}
+x_dp_{xd}^\prime 
+\frac{\lambda_p^2}{2\gamma_0^2} -\frac{\lambda_p^3}{2\gamma_0^2} 
+\Bigl(1-\lambda_p+\frac{x_d}{\rho_0}\Bigr)\frac{p_{xd}^2}{2}
-\frac{x_d}{\rho_0}\,\Bigl(\lambda_p -\frac{\lambda_p^2}{2\gamma_0^2} \Bigr)
+\frac{K_x}{2}\,x_d^2 +\frac{S_x}{6}\,x_d^3 \,.
\end{split}
\eq
The coefficients of the terms linear in $x_\beta$ and $p_{x\beta}$ vanish, as required.
The terms in the first and second rows describe the betatron motion and chromatic corrections, respectively.
The remaining terms describe the dispersion motion and are given by
\bq
\label{eq:kdmag}
\begin{split}
K_{d,\rm mag} &= \frac{\partial\tilde{G}_m}{\partial s}
+x_dp_{xd}^\prime 
+\frac{\lambda_p^2}{2\gamma_0^2} 
-\frac{\lambda_p^3}{2\gamma_0^2} 
+\frac{x_d^\prime p_{xd}^2}{2}
-\frac{x_d}{\rho_0}\,\Bigl(\lambda_p -\frac{\lambda_p^2}{2\gamma_0^2} \Bigr)
+\frac{K_x}{2}\,x_d^2 +\frac{S_x}{6}\,x_d^3 
\\
&\simeq
-(D_1 D_1^\pprime +D_1^{\prime2}) \,\frac{\lambda_p^2}{2} 
\\
&\quad
- \biggl[\, 
D_2 D_1^\pprime 
+2D_1 D_2^\pprime 
+3D_1^\prime D_2^\prime
+2D_1 D_1^\pprime 
+2D_1^{\prime2} 
-\frac{2D_1^2D_1^\pprime}{\rho_0}
-\frac{4D_1D_1^{\prime2}}{\rho_0}
-2D_1^2D_1^\prime\Bigl(\frac{1}{\rho_0}\Bigr)^\prime
\,\biggr]\, \frac{\lambda_p^3}{3} 
\\
&\quad
+\frac{\lambda_p^2}{2\gamma_0^2} 
-\frac{\lambda_p^3}{2\gamma_0^2} 
-\frac{D_1}{\rho_0}\,\lambda_p^2 
+\frac{K_x}{2}\,D_1^2 \lambda_p^2 
\\
&\quad
+D_1D_1^\pprime \,\lambda_p^2
+\biggl[\, D_2 D_1^\pprime 
+D_1 D_2^\pprime 
-D_1D_1^\pprime \Bigl(\frac{D_1}{\rho_0}-1\Bigr) 
-\frac{D_1D_1^{\prime2}}{\rho_0} \,\biggr] \,\lambda_p^3 
-D_1^2\,D_1^\prime \Bigl(\frac{1}{\rho_0}\Bigr)^\prime \lambda_p^3 
\\
&\quad
+\frac{D_1^{\prime2}}{2} \,\lambda_p^2
+\biggl[\, D_1^\prime D_2^\prime -\frac{D_1^{\prime2}}{2}\Bigl(\frac{D_1}{\rho_0} -1\Bigr) \,\biggr] \,\lambda_p^3 
\\
&\quad
-\frac{D_2}{\rho_0}\,\lambda_p^3
+\frac{D_1}{\rho_0}\,\frac{\lambda_p^3}{2\gamma_0^2} 
+K_x D_1D_2 \lambda_p^3
\\
&\quad
-D_1\, \biggl[\, D_2^\pprime +K_xD_2 +\frac{1}{2\gamma_0^2\rho_0} 
-\frac{D_1^{\prime2}}{2\rho_0} -\Bigl(\frac{D_1}{\rho_0}-1\Bigr)D_1^\pprime
\,\biggr]\,\frac{\lambda_p^3}{3}
+D_1^2D_1^\prime \Bigl(\frac{1}{\rho_0}\Bigr)^\prime \frac{\lambda_p^3}{3}
\\
&= \Bigl(\frac{1}{\gamma_0^2}-\frac{D_1}{\rho_0}\Bigr)\,\frac{\lambda_p^2}{2} 
-\biggl[\,\frac{D_2}{\rho_0}
-\frac{D_1}{\gamma_0^2\rho_0}
+\frac{D_1^{\prime2}}{2}
+\frac{3}{2\gamma_0^2} 
\,\biggr]\,\frac{\lambda_p^3}{3} \,.
\end{split}
\eq
The transformed Hamiltonian is, to the third order
\bq
\begin{split}
K_{\rm mag} &=
\Bigl(1+\frac{x_\beta}{\rho_0}\Bigr)\frac{p_{x\beta}^2}{2}
+\frac{K_x}{2}\,x_\beta^2 +\frac{S_x}{6}\,x_\beta^3 
\\
&\quad 
+\frac{p_{x\beta}^2}{2}\,\Bigl(\frac{x_d}{\rho_0} -\lambda_p\Bigr)
+\frac{x_\beta}{\rho_0}\,p_{x\beta}p_{xd}
+\frac{S_x}{2}\,x_\beta^2x_d
\\
&\quad 
+\Bigl(\frac{1}{\gamma_0^2}-\frac{D_1}{\rho_0}\Bigr)\,\frac{\lambda_p^2}{2} 
-\biggl[\,\frac{D_2}{\rho_0}
-\frac{D_1}{\gamma_0^2\rho_0}
+\frac{D_1^{\prime2}}{2}
+\frac{3}{2\gamma_0^2} 
\,\biggr]\,\frac{\lambda_p^3}{3} \,.
\end{split}
\eq
The differential time of flight is
\bq
\begin{split}
\frac{d\sigma}{ds} = \frac{\partial K_{\rm mag}}{\partial\lambda_p}
= \Bigl(\frac{1}{\gamma_0^2}-\frac{D_1}{\rho_0}\Bigr)\,\lambda_p 
-\biggl[\,\frac{D_2}{\rho_0}
-\frac{D_1}{\gamma_0^2\rho_0}
+\frac{D_1^{\prime2}}{2}
+\frac{3}{2\gamma_0^2} 
\,\biggr]\,\lambda_p^2 \,.
\end{split}
\eq
This matches the expression in eq.~\eqref{eq:tofallmag}, derived using geometry.

\subsection{Betatron and dispersion motion: all-electric ring}
We follow the same formal procedure for an all-electric ring.
The orbital bending and focusing comes entirely from the electrostatic potential.
The Hamiltonian is 
\bq
\begin{split}
K_e &= \lambda_p -\Bigl(1+\frac{x}{\rho_0}\Bigr)\sqrt{1 +2(\lambda_p-\Psi)+\beta_0^2(\lambda_p-\Psi)^2 - p_x^2} \,.
\\
&= \lambda_p 
-\Bigl(1+\frac{x}{\rho_0}\Bigr)\biggl[\, (1 +\lambda_p-\Psi)^2-\frac{(\lambda_p-\Psi)^2}{\gamma_0^2} - p_x^2 \,\biggr]^{1/2} \;.
\end{split}
\eq
We expand the square root to the third order
\bq
\begin{split}
K_e &\simeq \lambda_p -\Bigl(1+\frac{x}{\rho_0}\Bigr)
\biggl[\, 1 +\lambda_p-\Psi 
-\frac{(\lambda_p-\Psi)^2}{2\gamma_0^2(1 +\lambda_p-\Psi)}
-\frac{p_x^2}{2(1 +\lambda_p-\Psi)} \,\biggr]
\\
&\simeq \lambda_p -\Bigl(1+\frac{x}{\rho_0}\Bigr)
\biggl[\, 1 +\lambda_p-\Psi 
-\frac{(\lambda_p-\Psi)^2}{2\gamma_0^2}
+\frac{(\lambda_p-\Psi)^3}{2\gamma_0^2}
-(1 -\lambda_p+\Psi)\frac{p_x^2}{2} \,\biggr]
\\
&\simeq -1 + \Psi 
+\frac{(\lambda_p-\Psi)^2}{2\gamma_0^2}
-\frac{(\lambda_p-\Psi)^3}{2\gamma_0^2}
+\Bigl(1 -\lambda_p+\Psi +\frac{x}{\rho_0}\Bigr)\frac{p_x^2}{2} 
\\
&\quad -\frac{x}{\rho_0}
\biggl[\, 1 +\lambda_p-\Psi -\frac{(\lambda_p-\Psi)^2}{2\gamma_0^2} \,\biggr] 
\\
&\simeq -1 +\Bigl(1 -\lambda_p +\frac{2x}{\rho_0}\Bigr)\frac{p_x^2}{2} 
-\frac{x}{\rho_0}\,\lambda_p
+\Bigl(\frac{1}{\rho_0^2}+\frac{\Psi_{,xx}}{2}\Bigr)x^2 
+\Bigl(\frac{\Psi_{,xx}}{2\rho_0}+\frac{\Psi_{,xxx}}{6}\Bigr)x^3 
\\
&\quad 
+\frac{(\lambda_p-x/\rho_0 -\frac12\Psi_{,xx}x^2)^2}{2\gamma_0^2}
-\frac{(\lambda_p-x/\rho_0)^3}{2\gamma_0^2}
+\frac{x}{\rho_0}\frac{(\lambda_p-x/\rho_0)^2}{2\gamma_0^2} 
\\
&\simeq -1 +\Bigl(1 -\lambda_p +\frac{2x}{\rho_0}\Bigr)\frac{p_x^2}{2} 
+\Bigl(\frac{3-\beta_0^2}{\rho_0^2}+\Psi_{,xx}\Bigr)\,\frac{x^2}{2}
+\Bigl(\frac{6}{\gamma_0^2\rho_0^3} +(2-\beta_0^2)\frac{3\Psi_{,xx}}{\rho_0}+\Psi_{,xxx}\Bigr)\,\frac{x^3}{6} 
\\
&\quad 
+\frac{\lambda_p^2}{2\gamma_0^2}
-\frac{\lambda_p^3}{2\gamma_0^2}
-\frac{x}{\rho_0}\,\biggl[\, (2-\beta_0^2)\lambda_p -\frac{2\lambda_p^2}{\gamma_0^2} \,\biggr]
-\frac{1}{2\gamma_0^2}\Bigl(\frac{5}{\rho_0^2} +\Psi_{,xx}\Bigr)\, x^2 \lambda_p  \,.
\end{split}
\eq
It is convenient to define the elctrostatic quadrupole and sextupole gradients
\begin{subequations}
\begin{align}
K_x &= \frac{3-\beta_0^2}{\rho_0^2} +\Psi_{,xx} \,,
\\
S_x &= \frac{6}{\gamma_0^2\rho_0^3} +(2-\beta_0^2)\frac{3\Psi_{,xx}}{\rho_0} +\Psi_{,xxx} \,.
\end{align}
\end{subequations}
The Hamiltonian is then, dropping the constant term,
\bq
\begin{split}
K_e &\simeq \Bigl(1 -\lambda_p +\frac{2x}{\rho_0}\Bigr)\frac{p_x^2}{2} 
+\frac{K_x}{2}\,x^2
+\frac{S_x}{6}\,x^3
\\
&\quad 
+\frac{\lambda_p^2}{2\gamma_0^2}
-\frac{\lambda_p^3}{2\gamma_0^2}
-\frac{x}{\rho_0}\,\biggl[\, (2-\beta_0^2)\lambda_p -\frac{2\lambda_p^2}{\gamma_0^2} \,\biggr]
-\frac{1}{2\gamma_0^2}\Bigl(\frac{5}{\rho_0^2} +\Psi_{,xx}\Bigr)\, x^2 \lambda_p  \,.
\end{split}
\eq
Hamilton's equations for $x$ and $p_x$ are
\begin{subequations}
\begin{align}
\frac{dx}{ds} &= \Bigl(1-\lambda_p+\frac{2x}{\rho_0}\Bigr)\,p_x \,,
\\
\frac{dp_x}{ds} &= 
-K_x x
-\frac{S_x}{2}\,x^2 
-\frac{p_x^2}{\rho_0}
+\frac{2-\beta_0^2}{\rho_0} \,\lambda_p
-\frac{2\lambda_p^2}{\gamma_0^2\rho_0}
+\frac{1}{\gamma_0^2} \Bigl(\frac{5}{\rho_0^2}+\Psi_{,xx} \Bigr)\,\lambda_px \,.
\end{align}
\end{subequations}
The dispersion motion satisfies
\begin{subequations}
\begin{align}
x_d^\prime &= \Bigl(1-\lambda_p+\frac{2x_d}{\rho_0}\Bigr)\,p_{xd} \,,
\\
p_{xd}^\prime &= 
-K_x x_d
-\frac{S_x}{2}\,x_d^2 
-\frac{p_{xd}^2}{\rho_0}
+\frac{2-\beta_0^2}{\rho_0} \,\lambda_p
-\frac{2\lambda_p^2}{\gamma_0^2\rho_0}
+\frac{1}{\gamma_0^2} \Bigl(\frac{5}{\rho_0^2}+\Psi_{,xx} \Bigr)\,\lambda_px_d \,.
\end{align}
\end{subequations}
The second order differential equation is
\bq
\begin{split}
x_d^\pprime &= 
\frac{2x_d^\prime}{\rho_0}\,p_x
+\Bigl(1-\lambda_p+\frac{2x_d}{\rho_0}\Bigr)\,p_{xd}^\prime
+2x_dp_{xd}\,\Bigl(\frac{1}{\rho_0}\Bigr)^\prime
\\
&\simeq
\Bigl(1-\lambda_p+\frac{2x_d}{\rho_0}\Bigr)\,
\biggl[\,
\frac{p_{xd}^2}{\rho_0}
-K_x x_d
-\frac{S_x}{2}\,x_d^2
+\frac{2-\beta_0^2}{\rho_0} \,\lambda_p
-\frac{2\lambda_p^2}{\gamma_0^2\rho_0}
+\frac{1}{\gamma_0^2} \Bigl(\frac{5}{\rho_0^2}+\Psi_{,xx} \Bigr)\,\lambda_px_d 
\,\biggr]
\\
&\quad
+2x_dp_{xd}\,\Bigl(\frac{1}{\rho_0}\Bigr)^\prime \,.
\end{split}
\eq
The first order dispersion satisfies the equation
\bq
D_1^\pprime +K_xD_1 = \frac{2-\beta_0^2}{\rho_0} \,.
\eq
The second order dispersion satisfies the equation
\bq
\begin{split}
D_2^\pprime +K_xD_2 &= 
-\frac{2}{\gamma_0^2\rho_0}
+\frac{D_1^{\prime2}}{\rho_0}
-\frac{S_x}{2}\,D_1^2
+\Bigl(\frac{5}{\rho_0^2}+\Psi_{,xx} \Bigr)\,\frac{D_1}{\gamma_0^2}
+\Bigl(\frac{2D_1}{\rho_0}-1\Bigr)\,D_1^\pprime
+2D_1D_1^\prime\,\Bigl(\frac{1}{\rho_0}\Bigr)^\prime \,.
\end{split}
\eq
Note also that
\bq
\begin{split}
p_{xd}^\prime = 
\Bigl(x_d^\pprime -\frac{2x_d^\prime}{\rho_0}\,p_x -2x_dp_{xd}\,\Bigl(\frac{1}{\rho_0}\Bigr)^\prime \Bigr)
\Bigl(1-\lambda_p+\frac{2x_d}{\rho_0}\Bigr)^{-1} \,.
\end{split}
\eq
Hence to the third order
\bq
\begin{split}
x_dp_{xd}^\prime &\simeq
x_d\Bigl(x_d^\pprime -\frac{2x_d^\prime}{\rho_0}\,p_x -2x_dp_{xd}\,\Bigl(\frac{1}{\rho_0}\Bigr)^\prime \Bigr)
\Bigl(1-\lambda_p+\frac{2x_d}{\rho_0}\Bigr)^{-1} 
\\
&\simeq
x_d x_d^\pprime \Bigl(1+\lambda_p-\frac{2x_d}{\rho_0}\Bigr)
-\frac{2x_dx_d^{\prime2}}{\rho_0} 
-2x_d^2x_d^\prime\,\Bigl(\frac{1}{\rho_0}\Bigr)^\prime 
\\
&\simeq
D_1D_1^\pprime \lambda_p^2
+\biggl[\,
D_2D_1^\pprime 
+D_1D_2^\pprime 
-D_1D_1^\pprime \Bigl(\frac{2D_1}{\rho_0}-1\Bigr) 
-\frac{2D_1D_1^{\prime2}}{\rho_0}
\,\biggr]\,\lambda_p^3
-2D_1^2D_1^\prime\,\Bigl(\frac{1}{\rho_0}\Bigr)^\prime \lambda_p^3 \,.
\end{split}
\eq
Also to the third order
\bq
\begin{split}
x_d^\prime p_{xd} 
&\simeq x_d^{\prime2} \Bigl(1-\lambda_p+\frac{2x_d}{\rho_0}\Bigr)^{-1}
\\
&\simeq x_d^{\prime2} \Bigl(1+\lambda_p-\frac{2x_d}{\rho_0}\Bigr)
\\
&\simeq D_1^{\prime2} \lambda_p^2 + \biggl[\, 2D_1^\prime D_2^\prime - D_1^{\prime2} \Bigl(\frac{2D_1}{\rho_0}-1\Bigr) \biggr]\,\lambda_p^2 \,.
\end{split}
\eq
Next
\bq
\begin{split}
p_{xd} &\simeq x_d^\prime \, \Bigl( 1 +\lambda_p -\frac{2x_d}{\rho_0} \Bigr)
\\
&\simeq (D_1^\prime \lambda_p + D_2^\prime \lambda_p^2) \, \Bigl( 1 +\lambda_p -2\lambda_p\,\frac{D_1}{\rho_0} \Bigr)
\\
&\simeq D_1^\prime \lambda_p + D_2^\prime \lambda_p^2 + D_1^\prime \lambda_p^2 - \frac{2D_1D_1^\prime}{\rho_0} \lambda_p^2 \,.
\end{split}
\eq
Then to the required order
\bq
\begin{split}
x_d\,\frac{\partial p_{xd}}{\partial \lambda_p} &= (D_1 \lambda_p + D_2 \lambda_p^2)\,
\Bigl( D_1^\prime + 2D_2^\prime \lambda_p + 2D_1^\prime \lambda_p - \frac{4D_1D_1^\prime}{\rho_0} \lambda_p \Bigr)
\\
&\simeq  D_1 D_1^\prime \lambda_p + \biggl[\, D_2 D_1^\prime + 2D_1 D_2^\prime +2D_1 D_1^\prime -\frac{4D_1^2D_1^\prime}{\rho_0}\,\biggr]\, \lambda_p^2 \,.
\end{split}
\eq
Hence to the required order (see eqs.~\eqref{eq:gbeta} and \eqref{eq:deftildeg})
\bq
\begin{split}
\tilde{G}_e &= -\int_0^{\lambda_p} x_d\,\frac{\partial p_{xd}}{\partial \lambda_p}\,d\lambda_p 
\\
&= -D_1 D_1^\prime \,\frac{\lambda_p^2}{2} 
- \biggl[\, D_2 D_1^\prime + 2D_1 D_2^\prime +2D_1 D_1^\prime -\frac{4D_1^2D_1^\prime}{\rho_0}\,\biggr]\, \frac{\lambda_p^3}{3} \,.
\end{split}
\eq
This is the same as $\tilde{G}_m$ except the coefficient of the term in $D_1^2D_1^\prime/\rho_0$ is larger by a factor of two.
The transformed Hamiltonian is
\bq
\begin{split}
K_\beta &\simeq \frac{\partial\tilde{G}_e}{\partial s} -x_d^\prime p_{x\beta} +(x_\beta+x_d)p_{xd}^\prime 
\\
&\quad
+\frac{\lambda_p^2}{2\gamma_0^2} -\frac{\lambda_p^3}{2\gamma_0^2} 
+\Bigl(1-\lambda_p+2\frac{x_\beta+x_d}{\rho_0}\Bigr)\,\frac{(p_{x\beta}+p_{xd})^2}{2}
+\frac{K_x}{2}\,(x_\beta+x_d)^2
+\frac{S_x}{6}\,(x_\beta+x_d)^3
\\
&\quad
-\frac{x_\beta+x_d}{\rho_0}\,\biggl[\, (2-\beta_0^2)\lambda_p -\frac{2\lambda_p^2}{\gamma_0^2} \,\biggr]
-\frac{1}{2\gamma_0^2}\Bigl(\frac{5}{\rho_0^2} +\Psi_{,xx}\Bigr)\, (x_\beta+x_d)^2 \lambda_p  
\\
&\simeq
\Bigl(1+\frac{2x_\beta}{\rho_0}\Bigr)\,\frac{p_{x\beta}^2}{2}
+\frac{K_x}{2}\,x_\beta^2
+\frac{S_x}{6}\,x_\beta^3
\\
&\quad
+\frac{p_{x\beta}^2}{2}\,\Bigl(\frac{2x_d}{\rho_0} -\lambda_p\Bigr)
+\frac{2x_\beta}{\rho_0}\,p_{x\beta} p_{xd}
+\frac{S_x}{2}\,x_\beta^2 x_d
-\frac{1}{2\gamma_0^2}\Bigl(\frac{5}{\rho_0^2} +\Psi_{,xx}\Bigr)\, x_\beta^2 \lambda_p  
\\
&\quad
+p_{x\beta}\,\biggl[\, -x_d^\prime +\Bigl(1-\lambda_p+\frac{2x_d}{\rho_0}\Bigr)p_{xd} \,\biggr]
\\
&\quad
+x_\beta\,\biggl[\, p_{xd}^\prime 
+K_x x_d
+\frac{S_x}{2}\,x_d^2 
+\frac{p_{xd}^2}{\rho_0}
-\frac{2-\beta_0^2}{\rho_0} \,\lambda_p
+\frac{2\lambda_p^2}{\gamma_0^2\rho_0}
-\frac{1}{\gamma_0^2} \Bigl(\frac{5}{\rho_0^2}+\Psi_{,xx} \Bigr)\,\lambda_px_d 
\,\biggr]
\\
&\quad
+\frac{\partial\tilde{G}_e}{\partial s} +x_d p_{xd}^\prime 
+\frac{\lambda_p^2}{2\gamma_0^2} -\frac{\lambda_p^3}{2\gamma_0^2} 
+\Bigl(1-\lambda_p+\frac{2x_d}{\rho_0}\Bigr)\,\frac{p_{xd}^2}{2}
\\
&\quad
-\frac{x_d}{\rho_0}\,\biggl[\, (2-\beta_0^2)\lambda_p -\frac{2\lambda_p^2}{\gamma_0^2} \,\biggr]
-\frac{1}{2\gamma_0^2}\Bigl(\frac{5}{\rho_0^2} +\Psi_{,xx}\Bigr)\, x_d^2 \lambda_p  
+\frac{K_x}{2}\,x_d^2
+\frac{S_x}{6}\,x_d^3 \,.
\end{split}
\eq
The coefficients of the terms linear in $x_\beta$ and $p_{x\beta}$ vanish, as required.
The terms in the first and second rows describe the pure betatron motion and the chromatic corrections, respectively.
The terms which describe the dispersion motion are given by
\bq
\label{eq:kdelec}
\begin{split}
K_{d,\rm elec} &= 
\frac{\partial\tilde{G}_e}{\partial s} 
+\frac{\lambda_p^2}{2\gamma_0^2} 
-\frac{\lambda_p^3}{2\gamma_0^2} 
+x_d p_{xd}^\prime 
+\frac{x_d^\prime p_{xd}^2}{2}
\\
&\quad
-\frac{x_d}{\rho_0}\,\biggl[\, (2-\beta_0^2)\lambda_p -\frac{2\lambda_p^2}{\gamma_0^2} \,\biggr]
-\frac{1}{2\gamma_0^2}\Bigl(\frac{5}{\rho_0^2} +\Psi_{,xx}\Bigr)\, x_d^2 \lambda_p  
+\frac{K_x}{2}\,x_d^2
+\frac{S_x}{6}\,x_d^3 
\\
&\simeq
-(D_1 D_1^\pprime +D_1^{\prime2}) \,\frac{\lambda_p^2}{2} 
\\
&\quad
- \biggl[\, 
D_2 D_1^\pprime 
+2D_1 D_2^\pprime 
+3D_1^\prime D_2^\prime
+2D_1 D_1^\pprime 
+2D_1^{\prime2} 
-\frac{4D_1^2D_1^\pprime}{\rho_0}
-\frac{8D_1D_1^{\prime2}}{\rho_0}
-4D_1^2D_1^\prime\Bigl(\frac{1}{\rho_0}\Bigr)^\prime
\,\biggr]\, \frac{\lambda_p^3}{3} 
\\
&\quad
+\frac{\lambda_p^2}{2\gamma_0^2} 
-\frac{\lambda_p^3}{2\gamma_0^2} 
-(2-\beta_0^2)\frac{D_1}{\rho_0}\,\lambda_p^2
+\frac{K_x}{2}\,D_1^2 \lambda_p^2
\\
&\quad
+D_1D_1^\pprime \lambda_p^2
+\biggl[\,
D_2D_1^\pprime 
+D_1D_2^\pprime 
-D_1D_1^\pprime \Bigl(\frac{2D_1}{\rho_0}-1\Bigr) 
-\frac{2D_1D_1^{\prime2}}{\rho_0}
\,\biggr]\,\lambda_p^3
-2D_1^2D_1^\prime\,\Bigl(\frac{1}{\rho_0}\Bigr)^\prime \lambda_p^3 
\\
&\quad
+\frac{D_1^{\prime2}}{2} \lambda_p^2 
+ \biggl[\, D_1^\prime D_2^\prime - \frac{D_1^{\prime2}}{2} \Bigl(\frac{2D_1}{\rho_0}-1\Bigr) \biggr]\,\lambda_p^3 
\\
&\quad
-(2-\beta_0^2)\,\frac{D_2}{\rho_0}\, \lambda_p^3 
+K_xD_1D_2 \lambda_p^3
+\frac{2D_1}{\gamma_0^2\rho_0} \,\lambda_p^3
-\frac{D_1^2}{2\gamma_0^2}\Bigl(\frac{5}{\rho_0^2} +\Psi_{,xx}\Bigr)\, \lambda_p^3  
\\
&\quad
-D_1\,\biggl[\,
D_2^\pprime +K_xD_2 
+\frac{2}{\gamma_0^2\rho_0}
-\frac{D_1^{\prime2}}{\rho_0}
-\frac{1}{\gamma_0^2} \Bigl(\frac{5}{\rho_0^2}+\Psi_{,xx} \Bigr)\,D_1
\\
&\qquad\qquad\qquad\qquad\qquad\qquad
-\Bigl(\frac{2D_1}{\rho_0}-1\Bigr)\,D_1^\pprime
-2D_1D_1^\prime\,\Bigl(\frac{1}{\rho_0}\Bigr)^\prime 
\,\biggr]\,\frac{\lambda_p^3}{3} \,.
\\
&=
\Bigl(\frac{1}{\gamma_0^2}-(2-\beta_0^2)\frac{D_1}{\rho_0}\Bigr)\,\frac{\lambda_p^2}{2}
\\
&\quad
-\biggl[\,
(2-\beta_0^2)\frac{D_2}{\rho_0}
+\frac{D_1^{\prime2}}{2}
-\frac{4D_1}{\gamma_0^2\rho_0}
+\frac{5D_1^2}{2\gamma_0^2\rho_0^2} 
+\frac{\Psi_{,xx}D_1^2}{2\gamma_0^2} 
+\frac{3}{2\gamma_0^2} 
\,\biggr]\,\frac{\lambda_p^3}{3} \,.
\end{split}
\eq
The transformed Hamiltonian is, to the third order
\bq
\begin{split}
K_{\rm elec} &=
\Bigl(1+\frac{2x_\beta}{\rho_0}\Bigr)\,\frac{p_{x\beta}^2}{2}
+\frac{K_x}{2}\,x_\beta^2
+\frac{S_x}{6}\,x_\beta^3
\\
&\quad
+\frac{p_{x\beta}^2}{2}\,\Bigl(\frac{2x_d}{\rho_0} -\lambda_p\Bigr)
+\frac{2x_\beta}{\rho_0}\,p_{x\beta} p_{xd}
+\frac{S_x}{2}\,x_\beta^2 x_d
-\frac{1}{2\gamma_0^2}\Bigl(\frac{5}{\rho_0^2} +\Psi_{,xx}\Bigr)\, x_\beta^2 \lambda_p  
\\
&\quad
+\Bigl(\frac{1}{\gamma_0^2}-(2-\beta_0^2)\frac{D_1}{\rho_0}\Bigr)\,\frac{\lambda_p^2}{2}
\\
&\quad
-\biggl[\,
(2-\beta_0^2)\frac{D_2}{\rho_0}
+\frac{D_1^{\prime2}}{2}
-\frac{4D_1}{\gamma_0^2\rho_0}
+\biggl(\frac{5}{\rho_0^2} +\Psi_{,xx}\biggr)\,\frac{D_1^2}{2\gamma_0^2}
+\frac{3}{2\gamma_0^2} 
\,\biggr]\,\frac{\lambda_p^3}{3} \,.
\end{split}
\eq
The differential time of flight is
\bq
\begin{split}
\frac{d\sigma}{ds} = \frac{\partial K_{\rm elec}}{\partial\lambda_p}
&=\Bigl(\frac{1}{\gamma_0^2}-(2-\beta_0^2)\frac{D_1}{\rho_0}\Bigr)\,\lambda_p
\\
&\quad
-\biggl[\,
(2-\beta_0^2)\frac{D_2}{\rho_0}
+\frac{D_1^{\prime2}}{2}
-\frac{4D_1}{\gamma_0^2\rho_0}
+\biggl(\frac{5}{\rho_0^2} +\Psi_{,xx}\biggr)\,\frac{D_1^2}{2\gamma_0^2}
+\frac{3}{2\gamma_0^2} 
\,\biggr]\,\lambda_p^2 \,.
\end{split}
\eq
This matches the expression in eq.~\eqref{eq:tofallelec}, derived using geometry.

\subsection{Homogeneous all-electric ring}
The expression for the Hamiltonian simplifies for a homogeneous ring.
Then $1/\rho_0=1/r_0$ everywhere.
I treat all all-electric ring here.
Then $D_1^\prime =D_2^\prime =0$ and $x_p^\prime =p_{xd}=0$.
Hence from eq.~\eqref{eq:kdelec}
\bq
\begin{split}
K_{d,\rm elec} &= 
\frac{\lambda_p^2}{2\gamma_0^2} 
-\frac{\lambda_p^3}{2\gamma_0^2} 
-\frac{x_d}{r_0}\,\biggl[\, (2-\beta_0^2)\lambda_p -\frac{2\lambda_p^2}{\gamma_0^2} \,\biggr]
-\frac{1}{2\gamma_0^2}\Bigl(\frac{5}{r_0^2} +\Psi_{,xx}\Bigr)\, x_d^2 \lambda_p  
+\frac{K_x}{2}\,x_d^2
+\frac{S_x}{6}\,x_d^3 
\\
&\simeq
\frac{\lambda_p^2}{2\gamma_0^2} 
-\frac{\lambda_p^3}{2\gamma_0^2} 
+\biggl[\, \frac{K_x}{2} -\frac{2-\beta_0^2}{r_0}\,\biggr]\,D_1\lambda_p^2 
+\biggl[\, K_xD_1 -\frac{2-\beta_0^2}{r_0}\,\biggr]\,D_2\lambda_p^3
\\
&\quad
+\frac{2D_1}{\gamma_0^2r_0} \,\lambda_p^3
-\frac{D_1^2}{2\gamma_0^2}\Bigl(\frac{5}{r_0^2} +\Psi_{,xx}\Bigr)\, \lambda_p^3
+\biggl[\, \frac{1}{\gamma_0^2r_0^3} +(2-\beta_0^2)\frac{\Psi_{,xx}}{2r_0} +\frac{\Psi_{,xxx}}{6} \,\biggr]\,D_1^3 \lambda_p^3
\\
&= \biggl(\frac{1}{\gamma_0^2} -(2-\beta_0^2)\frac{D_1}{r_0}\biggr)\,\frac{\lambda_p^2}{2}
\\
&\quad
-\biggl[\, \frac{1}{2\gamma_0^2} 
-\frac{2D_1}{\gamma_0^2r_0} 
+\frac{D_1^2}{2\gamma_0^2}\Bigl(\frac{5}{r_0^2} +\Psi_{,xx}\Bigr)
-\biggl( \frac{1}{\gamma_0^2r_0^3} +(2-\beta_0^2)\frac{\Psi_{,xx}}{2r_0} +\frac{\Psi_{,xxx}}{6} \biggr)\,D_1^3 
\,\biggr]\,\lambda_p^3 \,.
\end{split}
\eq
Notice the term in $D_2$ cancels out, so we do not need to calculate it.
The first order dispersion is
\bq
D_1 = \frac{2-\beta_0^2}{2-n-\beta_0^2}\,r_0 \,.
\eq
Also for a homogeneous ring
\bq
\Psi_{,xx} = -\frac{n+1}{r_0^2} \,,\qquad
\Psi_{,xxx} = \frac{(n+1)(n+2)}{r_0^3} \,.
\eq
Hence
\bq
\frac{5}{r_0^2} +\Psi_{,xx} = \frac{4-n}{r_0^2} \,.
\eq
Also
\bq
\begin{split}
\frac{1}{\gamma_0^2r_0^3} +(2-\beta_0^2)\frac{\Psi_{,xx}}{2r_0} +\frac{\Psi_{,xxx}}{6} 
&= \frac{1}{\gamma_0^2r_0^3} -\Bigl(1+\frac{1}{\gamma_0^2}\Bigr)\frac{n+1}{2r_0^3} +\frac{(n+1)(n+2)}{6r_0^3} 
\\
&= \frac{1-n}{2\gamma_0^2r_0^3} +\frac{n^2-1}{6r_0^3} \,.
\end{split}
\eq
Then
\bq
\begin{split}
K_{d,\rm elec} &= \biggl(\frac{1}{\gamma_0^2} -(2-\beta_0^2)\frac{D_1}{r_0}\biggr)\,\frac{\lambda_p^2}{2}
\\
&\quad
+\frac{n^2-1}{6}\,\frac{D_1^3}{r_0^3}\,\lambda_p^3 
-\frac{1 -4(D_1/r_0) +(4-n)(D_1/r_0)^2 -(1-n)(D_1/r_0)^3}{2\gamma_0^2} \,\lambda_p^3 
\\
&= \biggl(\frac{1}{\gamma_0^2} -(2-\beta_0^2)\frac{D_1}{r_0}\biggr)\,\frac{\lambda_p^2}{2}
\\
&\quad
+\frac{n^2-1}{6}\,\frac{D_1^3}{r_0^3}\,\lambda_p^3 
-\frac{1}{2\gamma_0^2} \Bigl(1-\frac{D_1}{r_0}\Bigr)\Bigl(1-3\,\frac{D_1}{r_0}+(1-n)\frac{D_1^2}{r_0^2}\Bigr) \,\lambda_p^3 \,.
\end{split}
\eq
Hence the formal parameters are
\begin{subequations}
\begin{align}
\bar{a} &= (2-\beta_0^2)\frac{D_1}{r_0} -\frac{1}{\gamma_0^2} \,,
\\
\bar{b} &= (1-n^2)\,\frac{D_1^3}{r_0^3}
+\frac{3}{\gamma_0^2} \Bigl(1-\frac{D_1}{r_0}\Bigr)\Bigl(1-3\,\frac{D_1}{r_0}+(1-n)\frac{D_1^2}{r_0^2}\Bigr)  \,.
\end{align}
\end{subequations}
These expressions can be used to determine the values of $\langle \lambda_p \rangle$ and $\langle \lambda_p^2 \rangle$
which can then be employed in eq.~\eqref{eq:dadthetaoffenergyavg}.

\section{\label{sec:numerical} Numerical results}
\begin{figure}[!htb]
\centering
\includegraphics[width=0.95\textwidth]{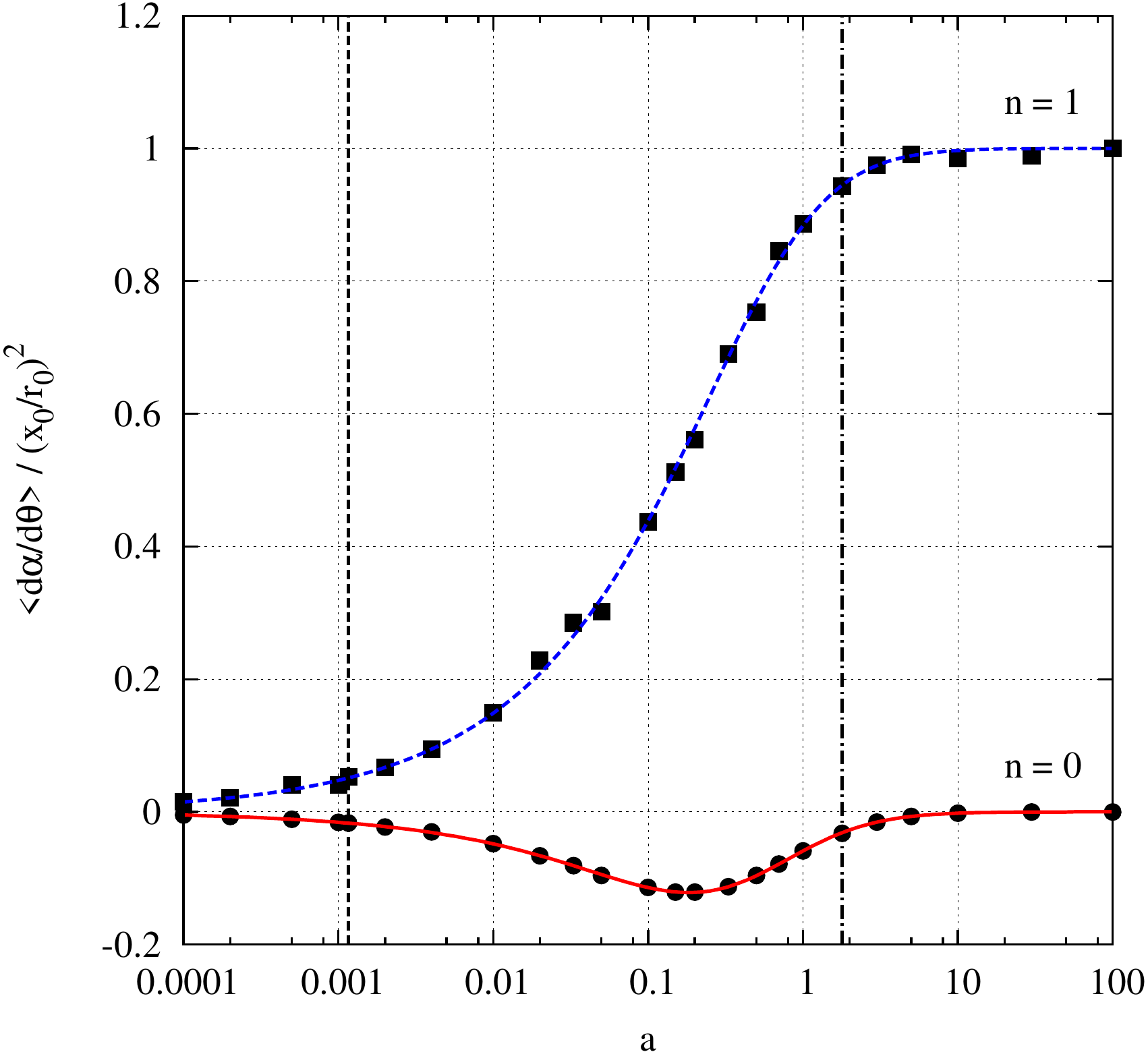}
\caption{\small
\label{fig:fig1}
Graph of $\langle d\alpha/d\theta\rangle / (x_0/r_0)^2$ {\em v}.~$a$.
The circles and squares are the tracking data for a field index of $n=0$ and 1,
viz.~a logarithmic potential and the Kepler problem, respectively.
The solid curves were plotted using analytical formulas derived in the text.
The dash and dotdash vertical lines denote the values of $a$ for a lepton and a proton, respectively.
}
\end{figure}

\begin{figure}[!htb]
\centering
\includegraphics[width=0.95\textwidth]{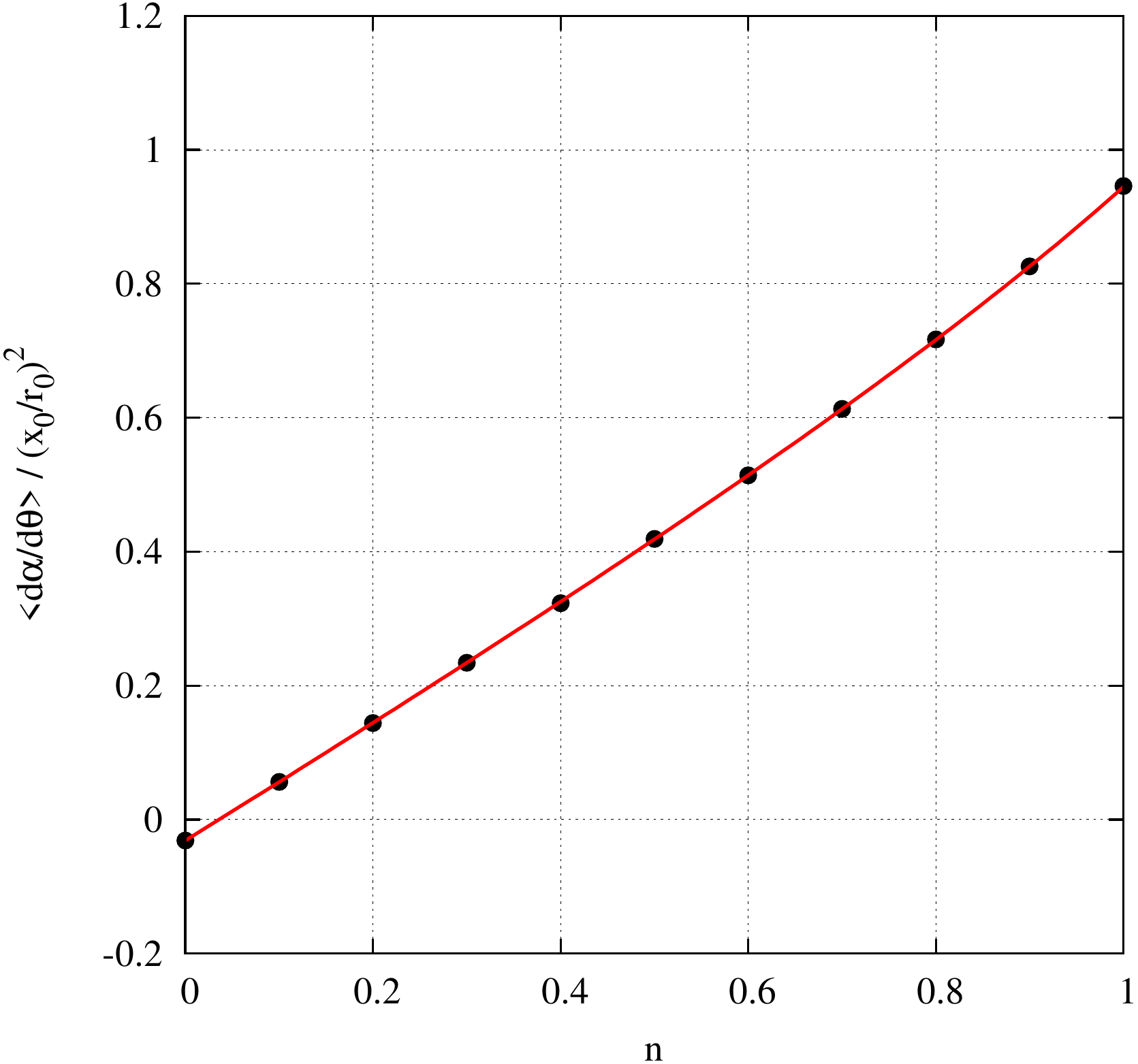}
\caption{\small
\label{fig:fig2}
Graph of $\langle d\alpha/d\theta\rangle / (x_0/r_0)^2$ {\em v}.~the field index $n$. 
The magnetic moment anomaly is that for a proton.
The circles are the tracking data and the solid curve was plotted using an analytical formula derived in the text.
}
\end{figure}

\begin{figure}[!htb]
\centering
\includegraphics[width=0.95\textwidth]{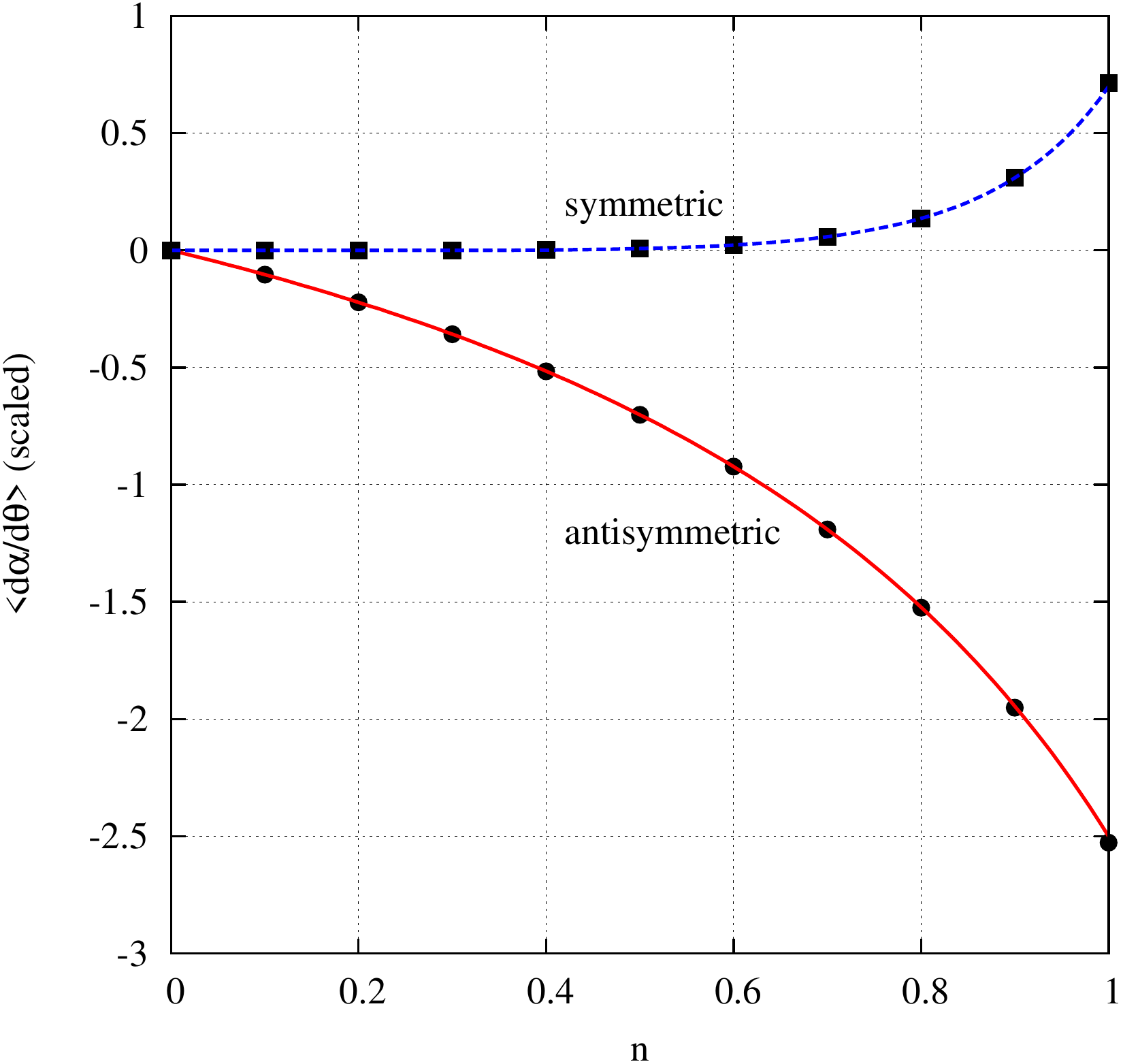}
\caption{\small
\label{fig:fig3}
Graph of the (scaled) symmetric and antisymmetric terms in $d\alpha/d\theta$ {\em v}.~the field index $n$, 
for off-energy (dispersion) orbits.
The squares and circles are the tracking data for the symmetric and antisymmetric terms
viz.~the parameters $\mathcal{S}$ and $\mathcal{A}$ defined in the text.
The solid and dashed curves were plotted using analytical formulas derived in the text.
}
\end{figure}

I employed a model homogeneous weak focusing all-electric ring with a radius of $r_0=40$ m
for all of the numerical simulations reported below.
In Figs.~\ref{fig:fig1} and \ref{fig:fig2}, I tracked betatron orbits
and in Fig.~\ref{fig:fig3} I tracked off-energy dispersion orbits.
In all cases, the particles were launched with their spins pointing along the reference orbit.

In both Figs.~\ref{fig:fig1} and \ref{fig:fig2}, 
I tracked one particle for one million turns, with an initial value $x_0=1$ mm and $p_{x0}=0$.
The energy was $H=H_0=\gamma_0 mc^2$.
A graph of $\langle d\alpha/d\theta\rangle/(x_0/r_0)^2$ {\em v}.~the magnetic moment anomaly $a$ 
is displayed in Fig.~\ref{fig:fig1}.
The circles and squares are the tracking data for a field index of $n=0$ and 1,
viz.~a logarithmic potential and the Kepler problem, respectively.
The solid curves were plotted using the formulas in eq.~\eqref{eq:dadt_n01}.
The dash and dotdash vertical lines denote the values of $a$ for a lepton and a proton, respectively.
The agreement between the tracking data and the analytical formula is excellent.
Next, in Fig.~\ref{fig:fig2}, 
I plotted a graph of $\langle d\alpha/d\theta\rangle / (x_0/r_0)^2$ {\em v}.~the field index $n$, for $0\le n \le 1$.
I set the magnetic moment anomaly to that for a proton, i.e.~$a\simeq1.792847$.
The circles are the tracking data and the solid curve was plotted using eq.~\eqref{eq:mydadt}.
The agreement between the tracking data and the analytical formula is excellent.
Note that the curve is almost a straight line, and the spin decoherence rate increases with the field index.

Next, I tracked the off-energy dispersion orbits.
In this case the orbits are circles and no averaging is necessary.
Since I derived the exact solution $d\alpha_d/d\theta$ in eq.~\eqref{eq:dadthetaoffenergy},
there was no need to use a small value for $\lambda_p$.
I set the magnetic moment anomaly to that for a proton and $\Delta H/H_0 = 0.1$.
Then $\lambda_p = (\Delta H/H_0)/\beta_0^2 \simeq 0.279$.
I tracked for ten thousand turns.
The point was simply to demonstrate that the tracking program was capable of accepting such a large value of $\Delta H/H_0$
and the numerically computed orbits remained circular to high accuracy.
(Typical results were $|r-r_d|/r_0 < O(10^{-13})$ and $|p_x|/p_0 < O(10^{-14})$.)
The initial conditions were $p_{x0}=0$ and $r=r_d$, 
computed using eq.~\eqref{eq:rdoffenergy}, by solving for $\gamma_d$ using
eqs.~\eqref{eq:gammadquadratic} and \eqref{eq:gammadkepler}.
I actually tracked two orbits, for $\pm\lambda_p$
and computed the following scaled symmetric and antisymmetric parameters
\begin{align}
\mathcal{S} = \frac{1}{2\lambda_p^2}\,\biggl\{\,\frac{d\alpha_d}{d\theta}\biggl|_{\lambda_p} 
+ \frac{d\alpha_d}{d\theta}\biggl|_{-\lambda_p} \,\biggr\} \,,
\\
\mathcal{A} = \frac{1}{2\lambda_p}\,\biggl\{\,\frac{d\alpha_d}{d\theta}\biggl|_{\lambda_p} 
- \frac{d\alpha_d}{d\theta}\biggl|_{-\lambda_p} \,\biggr\} \,.
\end{align}
The antisymmetric and symmetric terms are essentially the coefficients of $\lambda_p$ and $\lambda_p^2$
in a Taylor series expansion of $d\alpha_d/d\theta$ in powers of $\lambda_p$.
A graph of the (scaled) symmetric and antisymmetric terms in $d\alpha_d/d\theta$ {\em v}.~the field index $n$
is shown in Fig.~\ref{fig:fig3}.
The squares and circles are the tracking data for the symmetric and antisymmetric terms
viz.~$\mathcal{S}$ and $\mathcal{A}$, respectively.
The solid and dashed curves were calculated using eq.~\eqref{eq:dadthetaoffenergy}.
Note that in a real beam, where we expect $\langle \lambda_p\rangle = 0$,
the spin decoherence rate will be proportional to $\mathcal{S}$.
From Fig.~\ref{fig:fig3}, the value of $\mathcal{S}$ is very small for $0 \le n \lesssim 0.6$
and then rises steeply with increasing $n$.

\begin{figure}[!htb]
\centering
\includegraphics[width=0.95\textwidth]{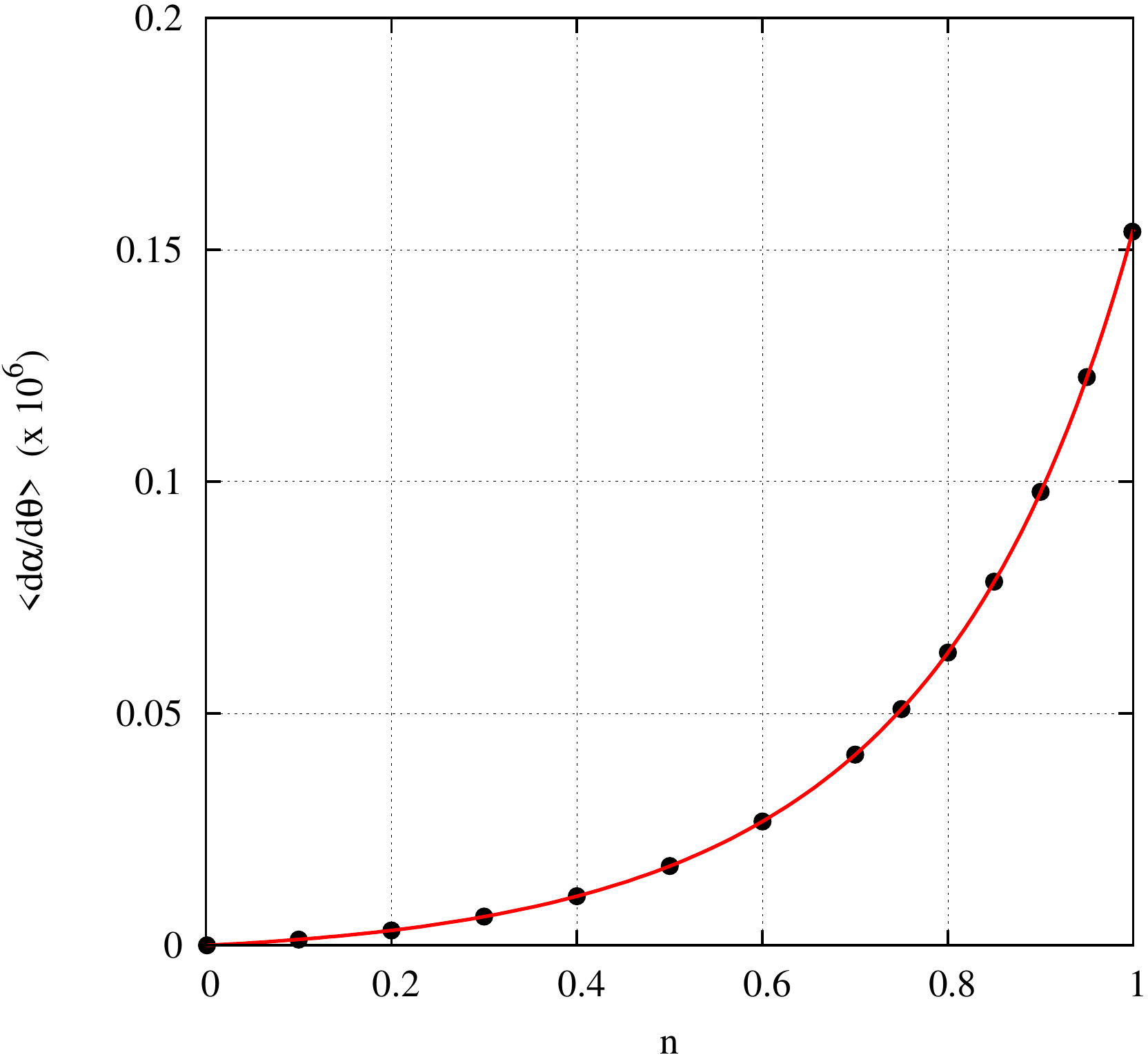}
\caption{\small
\label{fig:fig4}
Graph of $\langle d\alpha/d\theta\rangle$ {\em v}.~the field index $n$, for synchrotron oscillations.
The magnetic moment anomaly is that for a proton.
The circles are the tracking data and the solid curve was plotted using an analytical formula derived in the text.
}
\end{figure}

In Fig.~\ref{fig:fig4}, I display simulation results for the spin decoherence due to synchrotron oscillations.
The numerical tracking data are displayed as the circles in Fig.~\ref{fig:fig4}
and the solid line is the value of $\langle d\alpha_d/d\theta\rangle$, 
calculated using eq.~\eqref{eq:dadthetaoffenergyavg} with the averages derived in Section \ref{sec:syncosc}.
One rf cavity was inserted in the ring, with a voltage of $1$ MV and a harmonic number of $h=1$.
One particle was tracked with an initial energy offset given by $\lambda_{p0} = 2\times 10^{-4}$ and an initial time lag of zero.
The orbital motion consisted of synchrotron oscillations only.
There is excellent agreement between tracking and theory.

\section{\label{sec:conc} Conclusion}
I treated a model of relativistic charged particle motion in a horizontal plane in a central field of force.
The model was a homogeneous weak focusing all-electric storage ring, with an arbitrary field index $n\ge0$.
I derived the expression for the spin decoherence rate,
for both off-energy dispersion orbits and betatron oscillations.
For the off-energy motion, the dispersion orbits are circles and I solved the problem exactly.
For the betatron oscillations, I employed perturbation theory and canonical transformations.
I have calculated the spin precession exactly for the relativistic Kepler problem \cite{ManeRelKepler}.
I verified that it confirmed the solution from perturbation theory.
I displayed graphs of data from numerical tracking simulations, 
and demonstrated that the results matched well with the analytical calculations.
In particular, I found that for a proton, the spin decoherence rate due to betatron oscillations
increases approximately linearly with the field index $n$.
On the other hand, the spin decoherence rate due to the energy spread
is very small for $0 \le n \lesssim 0.6$ and then rises steeply with increasing $n$.

I reprodced, with permission, Ivan Koop's \cite{Koopprivcomm}
elegant analysis of motion in a vertical spiral in a logarithmic potential in Appendix \ref{sec:koopsoln}.
I also derived the exact solution for the orbit and spin motion in a vertical spiral in a logarithmic potential, in Appendix \ref{sec:logpot}.
Finally, a commentary on an analysis of the spin decoherence rate by Talman and Talman \cite{TalmanIPAC2012}
is presented in some Appendices below, including a discussion of some quantitative errors in their analysis.


\vfill\pagebreak
\appendix
\section{\label{sec:koopsoln} Analysis by Koop of vertical spiral motion in a radial electric field}
The following is reproduced with the kind permission of Ivan Koop \cite{Koopprivcomm}.
Koop solves for the particle kinetic energy and orbit radius for spiral motion in a logarithmic potential.
Note that Koop employs the synchronous condition and uses `$y$' to denote motion along the reference orbit
and denotes the pitch angle of the helix by $\theta$.
The synchronous condition means that the projection of the motion in the horizontal plane
circulates at the same frequency as the reference particle.
{\em (N.B.:~it is easily deduced that the synchronous condition requires $H \ne H_0$, i.e.~Koop's solution implies a nonzero energy offset.)}

\begin{quote}
Consider a particle which spirals upward with a small angle $\theta \approx v_z/v_y$
and still is synchronous with the RF frequency. That means:
\bq
\frac{v_y}{r} = \frac{v_0}{r_0} = \omega_0 \,.
\eq
Let us show that $r \le r_0$ and, hence, $v_y \le v_0$. 
Again, from the equality of the centripetal force and of the radial electric force one can deduce:
\bq
\frac{\gamma mvy^2}{r} = \frac{eE_0r_0}{r} \,.
\eq
Substituting into the above equation:
\bq
v_y^2(1+\theta^2) = 1 - \frac{1}{\gamma^2} \,,
\eq
one gets:
\bq
\gamma - \frac{1}{\gamma} = (1+\theta^2)\,\biggl(\gamma_0 - \frac{1}{\gamma_0}\biggr) \,.
\eq
Thus, the approximate value of the kinetic energy shift is equal to:
\bq
\frac{\Delta\gamma}{\gamma} = \theta^2\,\frac{\gamma_0^2-1}{\gamma_0^2+1} \ge 0 \,.
\eq
Now, from the equality $\gamma v_y = \gamma_0v_0^2$ one can deduce:
\bq
v_y = v_0\,\sqrt{\frac{\gamma_0}{\gamma}} \approx v_0\,\biggl(1 - \frac{\Delta\gamma}{2\gamma}\biggr)
= v_0\,\biggl(1 - \frac{\theta^2}{2}\,\frac{\gamma_0^2-1}{\gamma_0^2+1}\biggr) \,.
\eq
Same will be a relative change of the radius of an orbit:
\bq
r = r_0\,\frac{v_y}{v_0}
= r_0\,\sqrt{\frac{\gamma_0}{\gamma}} \approx r_0\,\biggl(1 - \frac{\Delta\gamma}{2\gamma}\biggr)
= r_0\,\biggl(1 - \frac{\theta^2}{2}\,\frac{\gamma_0^2-1}{\gamma_0^2+1}\biggr) \,.
\eq
\end{quote}

\vfill\pagebreak
\section{\label{sec:logpot} Logarithmic potential}
I solve for both the orbit and spin for motion in a vertical spiral in a logarithmic potential.
The field and potential are, for {\em all} orbits
\bq
\bm{E} = -\frac{E_0r_0}{r} \,,\qquad V = E_0r_0\,\ln\frac{r}{r_0} \,.
\eq
There is no vertical focusing.
The vertical momentum is a dynamical invariant $p_z=p_{z0}=\textrm{const}$.
Suppose the trajectory is a vertical spiral, of radius $r_*$, say,
hence there is no radial momentum, i.e.~$p_r=0$.
Let the pitch angle of the spiral be $\vartheta_*$.
Then $r_*$ and $\vartheta_*$ are both constants of the motion.
(Note that in general $r_*\ne r_0$; this will be discussed below.)
The centripetal force yields
\bq
\frac{m\gamma v_\theta^2}{r} = \frac{eE_0r_0}{r} \,.
\eq
Hence using $eE_0r_0 = mc^2 \gamma_0\beta_0^2$, 
we deduce $\gamma \beta_\theta^2 = \gamma_0\beta_0^2$.
Now by definition $\tan\vartheta_* = \beta_z/\beta_\theta$, so $\beta_\theta = \beta \cos\vartheta_*$.
Hence
\bq
\label{eq:equategm1g}
\gamma - \frac{1}{\gamma} = \gamma \beta^2 = \frac{\gamma_0\beta_0^2}{\cos^2\vartheta_*} \,.
\eq
{\em\color{red}
Hence the value of $\gamma$ is determined solely by the reference value $\gamma_0$ and the pitch angle $\vartheta_*$,
and does {\em not} depend on orbit radius $r_*$ or the total energy $H$.
}
Next
\bq
\gamma^2 - \gamma\,\frac{\gamma_0\beta_0^2}{\cos^2\vartheta_*} - 1 = 0 \,.
\eq
This can be solved exactly. We want the positive root, say $\gamma_*$,
\bq
\gamma_* = \frac12\,\biggl[\, \frac{\gamma_0\beta_0^2}{\cos^2\vartheta_*}
+\sqrt{\frac{\gamma_0^2\beta_0^4}{\cos^4\vartheta_*} + 4}\,\biggr] \,.
\eq
We can also solve approximately for $|\vartheta_*|\ll1$. 
Using eq.~\eqref{eq:equategm1g} and writing $\gamma_* = \gamma_0+\Delta\gamma_*$,
\begin{align}
\gamma_0-\frac{1}{\gamma_0} + \Bigl(1+\frac{1}{\gamma_0^2}\Bigr)\Delta\gamma_* &\simeq
\frac{\gamma_0\beta_0^2}{\cos^2\vartheta_*} \,,
\\
\gamma_0\beta_0^2 + \frac{\gamma_0^2+1}{\gamma_0^2}\,\Delta\gamma_* &\simeq
\frac{\gamma_0\beta_0^2}{\cos^2\vartheta_*} \,,
\\
(\gamma_0^2+1)\,\frac{\Delta\gamma_*}{\gamma_0} &\simeq \gamma_0^2\beta_0^2\,\tan^2\vartheta_* \,,
\\
\frac{\Delta\gamma_*}{\gamma_0} &\simeq \vartheta_*^2\, \frac{\gamma_0^2-1}{\gamma_0^2+1} \,.
\end{align}
We determine the radius from the total energy $H$.
We do {\em not} need to constrain the energy to be $H_0$;
we can permit an offset $\Delta H/H_0 = \beta_0^2\,\lambda_p$.
Then, using $H = \gamma mc^2 + e\Phi$,
\begin{align}
H &= \gamma_* mc^2 + eE_0r_0\,\ln\frac{r_*}{r_0} \,,
\\
\gamma_0(1+\beta_0^2\lambda_p) &= \gamma_0 +\Delta\gamma_* + \gamma_0\beta_0^2\,\ln\frac{r_*}{r_0} \,,
\\
\lambda_p - \frac{\Delta\gamma_*}{\gamma_0\beta_0^2} &= \ln\frac{r_*}{r_0} \,,
\\
r_* &= r_0\,e^{\lambda_p - \Delta\gamma_*/(\gamma_0\beta_0^2)} \,.
\end{align}
This is the exact solution.
Note that in general $r_*\ne r_0$.
Next we treat the spin motion and the helicity. Recall that
\bq
\frac{d\ }{dt}(\bm{s}\cdot\hat{\bm{\beta}}) = \frac{e}{mc}\biggl(a - \frac{1}{\beta^2\gamma^2}\biggr)
(\bm{\beta}\times\bm{E})\cdot(\bm{s}\times\hat{\bm{\beta}})\,.
\eq
In this model $\bm{E} \parallel \hat{\bm{r}}$ and $\bm{\beta}$ has no radial component, so $\bm{E}\cdot\bm{\beta}=0$, so
\bq
(\bm{\beta}\times\bm{E})\cdot(\bm{s}\times\hat{\bm{\beta}})
= \bm{\beta}\cdot\bm{s}\,\bm{E}\cdot\hat{\bm{\beta}} - \bm{\beta}\cdot\hat{\bm{\beta}}\,\bm{E}\cdot\bm{s}
= -E\beta_*\,\bm{s}\cdot\hat{\bm{r}}
\eq
Hence using $d\theta/dt = c\beta_\theta/r_*$,
\bq
\label{eq:evolhelicitylogpot}
\begin{split}
\frac{d\ }{d\theta}(\bm{s}\cdot\hat{\bm{\beta}}) 
&= -\frac{r_*}{c\beta_\theta}\,\frac{eE\beta_*}{mc}\biggl(a - \frac{1}{\beta_*^2\gamma_*^2}\biggr)\,\bm{s}\cdot\hat{\bm{r}}
\\
&= \frac{eE_0r_0}{mc^2 \cos\vartheta_*}\biggl(a - \frac{1}{\beta_*^2\gamma_*^2}\biggr)\,\bm{s}\cdot\hat{\bm{r}}
\\
&= \frac{\gamma_0\beta_0^2}{\cos\vartheta_*}\,\biggl(a - \frac{1}{\beta_*^2\gamma_*^2}\biggr)\,\bm{s}\cdot\hat{\bm{r}}
\\
&= \frac{1}{\gamma_0\cos\vartheta_*}\,\biggl(1 - \frac{\beta_0^2\gamma_0^2}{\beta_*^2\gamma_*^2}\biggr)\,\bm{s}\cdot\hat{\bm{r}} \,.
\end{split}
\eq
In the last line, the value of $a$ was set using the magic gamma $a=1/(\beta_0^2\gamma_0^2)$.
Notice that $\bm{s}\cdot\hat{\bm{r}}$ appears on the right-hand side of the above equation.
Hence we cannot simply write $\bm{s}\cdot\hat{\bm{\beta}} = \cos\alpha$
and derive an equation purely for $d\alpha/d\theta$ as was done for case of motion in the horizontal plane.
We must solve for the spin motion explicitly.
We do so as follows.
In the absence of a magnetic field, the spin precession equation of motion is
\bq
\frac{d\bm{s}}{dt} = \bm{\Omega}\times\bm{s}
= \frac{e}{mc}\,\biggl(a + \frac{1}{\gamma+1}\biggr)\,(\bm{\beta}\times\bm{E}) \times \bm{s} \,.
\eq
The spin precession vector $\bm{\Omega}$ points along $\bm{\beta} \times \bm{E}$.
Recall that $\bm{E}$ points radially and $\bm{\beta}$ points along the tangent to the orbit helix,
hence in a cylindrical polar coordinate system, 
$\bm{\beta} \times \bm{E}$ has fixed (in time) components along $\hat{\bm{\theta}}$ and $\hat{\bm{z}}$, viz.
\bq
\bm{\beta} \times \bm{E} \; \parallel \;
\bm{\beta} \times \hat{\bm{r}} = (\beta_\theta\,\hat{\bm{\theta}} + \beta_z\,\hat{\bm{z}})\times\hat{\bm{r}}
= -\beta_\theta\,\hat{\bm{z}} + \beta_z\,\hat{\bm{\theta}} \,.
\eq
Next we need to express the evolution of the spin components in cylindrical polar coordinates
\bq
\begin{split}
\frac{d\bm{s}}{dt} &= \frac{ds_r}{dt}\,\hat{\bm{r}} +\frac{ds_\theta}{dt}\,\hat{\bm{\theta}} +\frac{ds_z}{dt}\,\hat{\bm{z}}
+s_r\,\frac{d\hat{\bm{r}}}{dt} +s_\theta\,\frac{d\hat{\bm{\theta}}}{dt}
\\
&= \frac{ds_r}{dt}\,\hat{\bm{r}} +\frac{ds_\theta}{dt}\,\hat{\bm{\theta}} +\frac{ds_z}{dt}\,\hat{\bm{z}}
+\frac{v_\theta}{r}\,\hat{\bm{z}} \times \bm{s} \,.
\end{split}
\eq
Then we derive an `effective' spin precession vector as follows
\bq
\frac{ds_r}{d\theta}\,\hat{\bm{r}} +\frac{ds_\theta}{d\theta}\,\hat{\bm{\theta}} +\frac{ds_z}{d\theta}\,\hat{\bm{z}}
= \Bigl(\frac{r}{v_\theta}\, \bm{\Omega} -\hat{\bm{z}} \Bigr) \times \bm{s}
\equiv \bar{\bm{\Omega}} \times \bm{s} \,.
\eq
In the language of General Relativity, this is a covariant derivative;
we have subtracted contribution from the time variation of the basis vectors $\hat{\bm{r}}$ and $\hat{\bm{\theta}}$.
Then 
\bq
\label{eq:baromega1}
\begin{split}
\bar{\bm{\Omega}} &= \frac{r}{v_\theta}\,\bm{\Omega} - \hat{\bm{z}}
\\
&= \frac{er_*}{mc^2\beta_\theta}\,\biggl(a + \frac{1}{\gamma+1}\biggr)\,(\bm{\beta}\times\bm{E}) 
-\hat{\bm{z}} 
\\
&= -\frac{eE_0r_0}{mc^2\beta_\theta}\,\biggl(a + \frac{1}{\gamma_*+1}\biggr)\,(-\beta_\theta\,\hat{\bm{z}} + \beta_z\,\hat{\bm{\theta}}) 
-\hat{\bm{z}} 
\\
&= \gamma_0\beta_0^2\,\biggl[\biggl(a +\frac{1}{\gamma_*+1} -\frac{1}{\gamma_0\beta_0^2}\biggr) \,\hat{\bm{z}}
-\biggl(a + \frac{1}{\gamma_*+1}\biggr) \frac{\beta_z}{\beta_\theta}\, \hat{\bm{\theta}} \biggr] \,.
\end{split}
\eq
This has the significant feature that the components of $\bar{\bm{\Omega}}$ 
are constant in time (or $\theta$), when expressed in cylindrical polar coordinates.
Hence the spin components $(s_r,s_\theta,s_z)$ precess in a plane
normal to $\bar{\bm{\Omega}}$, at angular frequency $|\bar{\bm{\Omega}}|$.
(Note that in the lab frame, the `plane' itself rotates around the vertical axis
at the angular frequency $v_\theta/r_*$.)
{\em\color{red} Note also that the orientation of this plane is {\em not} equal to the pitch angle $\vartheta_*$ of the orbital motion.}
Unlike the case of motion in the horizontal plane, the spin does {\em not} precess in the same plane as the orbit.
Let $\bar{\bm{\Omega}}$ be oriented at an angle $\varphi_*$ to the horizontal.
Then define $\bar{\Omega}$ via
\bq
\bar{\bm{\Omega}} = \bar{\Omega}\,(\cos\varphi_*\,\hat{\bm{\theta}} + \sin\varphi_*\,\hat{\bm{z}}) \,.
\eq
Then
\bq
\tan\varphi_* = -\frac{a +1/(\gamma_*+1) -1/(\gamma_0\beta_0^2)}{a + 1/(\gamma_*+1)} \,\frac{\beta_\theta}{\beta_z} \,.
\eq
Let us define a right-handed orthonormal basis $(\bm{\zeta}_1,\bm{\zeta}_2,\bm{\zeta}_3)$ as follows
\begin{subequations}
\begin{align}
\bm{\zeta}_1 &= \phantom{-} \hat{\bm{r}} \,,
\\
\bm{\zeta}_2 &= \phantom{-} \cos\varphi_*\,\hat{\bm{\theta}} + \sin\varphi_*\,\hat{\bm{z}} \,,
\\
\bm{\zeta}_3 &= -\sin\varphi_*\,\hat{\bm{\theta}} + \cos\varphi_*\,\hat{\bm{z}} \,.
\end{align}
\end{subequations}
Then $\bm{\zeta}_1 \parallel \bm{E}$,
$\bm{\zeta}_2 \parallel \bar{\bm{\Omega}}$ and 
$\bm{\zeta}_3 = \bm{\zeta}_1 \times \bm{\zeta}_2$.
Our interest is in the case when the spin points initially along the reference orbit
\bq
\bm{s}(0) = \hat{\bm{\theta}} = \cos\varphi_*\,\bm{\zeta}_2 -\sin\varphi_*\,\bm{\zeta}_3 \,.
\eq
Since $\bar{\bm{\Omega}}$ has constant (in $\theta$) components in cylindrical polar coordinates,
the component of $\bm{s}$ parallel to $\bar{\bm{\Omega}}$ is invariant,
and the other components of $\bm{s}$ precess in the $(\bm{\zeta}_1,\bm{\zeta}_3)$ plane
at the angular frequency $\bar{\Omega}$.
Hence  
\bq
\bm{s} = \cos\varphi_*\,\bm{\zeta}_2 
-\sin\varphi_*\,\bigl[\, \cos(\bar{\Omega}\theta) \bm{\zeta}_3 +\sin(\bar{\Omega}\theta) \bm{\zeta}_1 \,\bigr] \,.
\eq
Then
\bq
\label{eq:sdotr}
\bm{s}\cdot\hat{\bm{r}} = \bm{s}\cdot\bm{\zeta}_1 = -\sin\varphi_*\,\sin(\bar{\Omega}\theta) \,.
\eq
We can now use this in eq.~\eqref{eq:evolhelicitylogpot}.
However, since we know $\bm{s}$ as a function of $\theta$,
let us solve for the evolution of the helicity directly.
Using $\hat{\bm{\beta}} = \cos\vartheta_*\,\hat{\bm{\theta}} + \sin\vartheta_*\,\hat{\bm{z}}$, the helicity is
\bq
\begin{split}
\bm{s}\cdot\hat{\bm{\beta}} &= 
\cos\varphi_*\, \bm{\zeta}_2 \cdot\hat{\bm{\beta}}
-\sin\varphi_*\, \cos(\bar{\Omega}\theta) \, \bm{\zeta}_3 \cdot\hat{\bm{\beta}} 
\\
&= \cos\varphi_*\,\cos(\varphi_*-\vartheta_*)
+\sin\varphi_*\,\sin(\varphi_*-\vartheta_*) \,\cos(\bar{\Omega}\theta) \,.
\end{split}
\eq
Hence 
\bq
\frac{d\ }{d\theta}(\bm{s}\cdot\hat{\bm{\beta}}) 
= -\bar{\Omega}\,\sin\varphi_*\,\sin(\varphi_*-\vartheta_*) \,\sin(\bar{\Omega}\theta) \,.
\eq
Note also, using $\bm{s}\cdot\hat{\bm{\beta}} = \cos\alpha$ that
$d(\bm{s}\cdot\hat{\bm{\beta}})/d\theta = -\sin\alpha\,(d\alpha/d\theta)$.
Hence
\bq
\frac{d\alpha}{d\theta} = \bar{\Omega}\, 
\frac{\sin\varphi_*\,\sin(\varphi_*-\vartheta_*)\, \sin(\bar{\Omega}\theta)}{\sin\alpha} \,.
\eq
We need an expression for $\sin\alpha$,
but only have an expression for $\cos\alpha = \bm{s}\cdot\hat{\bm{\beta}}$.
Using $\sin^2\alpha = 1 - \cos^2\alpha = 1 - (\bm{s}\cdot\hat{\bm{\beta}})^2$,
we {\em provisionally} write
\bq
\frac{d\alpha}{d\theta} = \bar{\Omega}\, 
\frac{\sin\varphi_*\,\sin(\varphi_*-\vartheta_*)\, \sin(\bar{\Omega}\theta)}{\sqrt{1 - (\bm{s}\cdot\hat{\bm{\beta}})^2}} \,.
\eq
The provision is that $\sin\alpha$ changes sign according as $\alpha$ is positive or negative, 
but the square root in the denominator is always positive.
Hence we must exercise some care to average over the orbit.
(Note also that because the angular velocity is constant, an average over $t$ is equivalent to an average over $\theta$.)
Hence define $\tilde{\phi} = \bar{\Omega}\theta$ and
\bq
\chi = \frac{\sin\varphi_*\,\sin(\varphi_*-\vartheta_*)\, \sin\tilde{\phi}}{\sqrt{1 - (\cos\varphi_*\,\cos(\varphi_*-\vartheta_*) 
+\sin\varphi_*\,\sin(\varphi_*-\vartheta_*)\, \cos\tilde{\phi} )^2 }} \,.
\eq
If we average this over a full period of $\tilde{\phi}$, the average will vanish
because the numerator is antisymmetric and the denominator is symmetric
under a transformation $\tilde{\phi} \to -\tilde{\phi}$.
We obtain the correct answer by averaging over half a period $0\le \tilde{\phi} \le \pi$
\bq
\langle\chi\rangle = \frac{1}{\pi}\,\int_0^\pi \chi \,d\tilde{\phi} \,.
\eq
We employ the substitution
\bq
u = \cos\varphi_*\,\cos(\varphi_*-\vartheta_*) 
+\sin\varphi_*\,\sin(\varphi_*-\vartheta_*)\, \cos\tilde{\phi}
\eq
The limits on $u$ are, for $\tilde{\phi} = 0$ and $\pi$ respectively,
\begin{subequations}
\begin{align}
u_0 &= 
\cos\varphi_*\,\cos(\varphi_*-\vartheta_*) +\sin\varphi_*\,\sin(\varphi_*-\vartheta_*)
= \cos\vartheta_* \,,
\\
u_\pi &= 
\cos\varphi_*\,\cos(\varphi_*-\vartheta_*) -\sin\varphi_*\,\sin(\varphi_*-\vartheta_*)
= \cos(2\varphi_*-\vartheta_*) \,.
\end{align}
\end{subequations}
Then
\bq
\label{eq:chiavg}
\begin{split}
\langle\chi\rangle &= -\frac{1}{\pi}\,
\int_{u_0}^{u_\pi} \frac{du}{\sqrt{1 - u^2}}
\\
&= \frac{1}{\pi}\,\biggl[\, \cos^{-1}u \,\biggr]_{u_0}^{u_\pi} 
\\
&= \frac{1}{\pi}\,
\biggl[\,(2\varphi_*-\vartheta_*) -\vartheta_* \,\biggr] 
\\
&= \frac{2(\varphi_*-\vartheta_*)}{\pi} \,.
\end{split}
\eq
Hence
\bq
\biggl\langle\frac{d\alpha}{d\theta}\biggr\rangle =
\frac{2(\varphi_*-\vartheta_*)}{\pi} \, \bar{\Omega} \,.
\eq
This is a formally exact solution; 
all quantities in the above expression are known exactly, in principle.
Note also from eq.~\eqref{eq:baromega1} that both $\bar{\Omega}$ and $\varphi_*$
depend {\em only} on the reference value $\gamma_0$ and the pitch angle of the helix $\vartheta_*$.
Hence we do {\em not} require knowledge of the orbit radius,
nor the total energy, to derive the above solution.

Let us now evaluate $\bar{\Omega}$ and $\varphi_*$ approximately, for $|\vartheta_*|\ll1$.
Note that by construction, $\bar{\bm{\Omega}}$ vanishes on the reference orbit.
Hence off-axis, $\bar{\bm{\Omega}}$ is a small quantity.
We determine the components of $\bar{\bm{\Omega}}$ as follows.
First, using $a=1/(\gamma_0^2-1)$,
\bq
\begin{split}
a + \frac{1}{\gamma_*+1} -\frac{1}{\gamma_0\beta_0^2}
&= \frac{1}{\gamma_*+1} - \frac{\gamma_0-1}{\gamma_0^2-1} 
\\
&= \frac{1}{\gamma_*+1} - \frac{1}{\gamma_0+1} 
\\
&= \frac{\gamma_0-\gamma_*}{(\gamma_*+1)(\gamma_0+1)} 
\\
&\simeq -\frac{\gamma_0}{(\gamma_0+1)^2} \,\frac{\Delta\gamma_*}{\gamma_0}
\\
&\simeq -\vartheta_*^2\,\frac{\gamma_0}{(\gamma_0+1)^2} \,\frac{\gamma_0^2-1}{\gamma_0^2+1}
\\
&\simeq -\vartheta_*^2\,\frac{\gamma_0}{\gamma_0^2+1} \,\frac{\gamma_0-1}{\gamma_0+1} \,.
\end{split}
\eq
Next
\bq
a + \frac{1}{\gamma_*+1} 
\simeq \frac{1}{\beta_0^2\gamma_0^2} + \frac{1}{\gamma_0+1} 
= \frac{1}{\gamma_0^2-1} + \frac{\gamma_0-1}{\gamma_0^2-1} 
= \frac{\gamma_0}{\gamma_0^2-1} 
= \frac{1}{\gamma_0\beta_0^2} \,. 
\eq
Substituting in eq.~\eqref{eq:baromega1},
\bq
\label{eq:baromega2}
\begin{split}
\bar{\bm{\Omega}} &= \gamma_0\beta_0^2\,
\biggl[\biggl(a +\frac{1}{\gamma_*+1} -\frac{1}{\gamma_*\beta_\theta^2} \biggr) \hat{\bm{z}}
-\biggl(a + \frac{1}{\gamma_*+1}\biggr) \frac{\beta_z}{\beta_\theta} \hat{\bm{\theta}} \biggr] 
\\
&\simeq -\gamma_0\beta_0^2\,
\biggl[\, \vartheta_*^2\,\frac{\gamma_0}{\gamma_0^2+1} \,\frac{\gamma_0-1}{\gamma_0+1} \,\hat{\bm{z}}
+\frac{1}{\gamma_0\beta_0^2} \,\vartheta_*\, \hat{\bm{\theta}} \,\biggr] 
\\
&\simeq -\vartheta_*^2\,\frac{(\gamma_0-1)^2}{\gamma_0^2+1} \, \hat{\bm{z}} - \vartheta_*\, \hat{\bm{\theta}} \,.
\end{split}
\eq
The horizontal and vertical components of $\bar{\bm{\Omega}}$ are of the first and second order in small quantities, respectively.
Hence for a small orbital pitch angle $|\vartheta_*|\ll1$, $\bar{\bm{\Omega}}$ is almost {\em horizontal}.
This demonstrates that the orientation of the off-axis components of the spin precession vector can be counterintuitive, 
for the various modes of off-axis orbital motion.
For $|\vartheta_*|\ll1$,
\begin{align}
\bar{\Omega} &\simeq -\vartheta_* \,,
\\
\varphi_* \simeq \tan\varphi_* &\simeq \frac{(\gamma_0-1)^2}{\gamma_0^2+1} \,\vartheta_* \,,
\\
\varphi_*-\vartheta_* &\simeq -\frac{2\gamma_0}{\gamma_0^2+1} \,\vartheta_* \,.
\end{align}
We now approximate for small $|\vartheta_*|$.
\bq
\label{eq:vertsctavg}
\biggl\langle\frac{d\alpha}{d\theta}\biggr\rangle \simeq
\frac{4\vartheta_*^2}{\pi}\, \frac{\gamma_0}{\gamma_0^2+1} \,.
\eq
As stated above, this depends only on the reference value $\gamma_0$ and the pitch angle of the helix $\vartheta_*$, 
but not on the orbit radius $r_*$ or the total energy $H$.

For the sake of completeness, let us derive the above result
by following the formal procedure using eq.~\eqref{eq:evolhelicitylogpot}.
Using eq.~\eqref{eq:sdotr},
\bq
\begin{split}
\frac{d\alpha}{d\theta} &=
\frac{1}{\gamma_0\cos\vartheta_*}\,
\biggl(1 - \frac{\beta_0^2\gamma_0^2}{\beta_*^2\gamma_*^2}\biggr)\,
\frac{\sin\varphi_*\,\sin(\bar{\Omega}\theta)}{\sqrt{1-(\bm{s}\cdot\hat{\bm{\beta}})^2}}
\\
&= \frac{1}{\gamma_0\cos\vartheta_*\sin(\varphi_*-\vartheta_*)}\,
\biggl(1 - \frac{\beta_0^2\gamma_0^2}{\beta_*^2\gamma_*^2}\biggr)\,\chi \,.
\end{split}
\eq
Then use eq.~\eqref{eq:chiavg} to deduce the average
\bq
\biggl\langle\frac{d\alpha}{d\theta} \biggr\rangle
= \frac{2}{\pi\gamma_0\cos\vartheta_*}\,
\biggl(1 - \frac{\beta_0^2\gamma_0^2}{\beta_*^2\gamma_*^2}\biggr)\, 
\frac{\varphi_*-\vartheta_*}{\sin(\varphi_*-\vartheta_*)} \,.
\eq
This is also a formally exact solution.
Next we need
\bq
1 - \frac{\beta_0^2\gamma_0^2}{\beta_*^2\gamma_*^2}
= 1 - \frac{\gamma_0^2-1}{\gamma_*^2-1}
\simeq \frac{\gamma_*^2-\gamma_0^2}{\beta_0^2\gamma_0^2}
\simeq \frac{(1+\Delta\gamma_*/\gamma_0)^2-1}{\beta_0^2}
\simeq \frac{2}{\beta_0^2}\,\frac{\Delta\gamma_*}{\gamma_0} \,.
\eq
Then for $|\vartheta_*|\ll1$,
\bq
\begin{split}
\biggl\langle\frac{d\alpha}{d\theta} \biggr\rangle
&\simeq \frac{4\vartheta_*^2}{\pi\gamma_0\beta_0^2\cos\vartheta_*}\,
\frac{\gamma_0^2-1}{\gamma_0^2+1}\,
\frac{\varphi_*-\vartheta_*}{\sin(\varphi_*-\vartheta_*)} 
\\
&\simeq \frac{4\vartheta_*^2}{\pi}\,\frac{\gamma_0}{\gamma_0^2+1} \,.
\end{split}
\eq
This agrees with eq.~\eqref{eq:vertsctavg}.

\vfill\pagebreak
\section{Spin decoherence rate}
Talman and Talman \cite{TalmanIPAC2012}
published the following formula for the spin decoherence rate 
for orbital and spin motion in the horizontal plane in an all-electric ring
(eq.~(17) in \cite{TalmanIPAC2012})
\bq
\label{eq:talmanipac2012_eq17}
-\biggl\langle \frac{d\alpha}{d\theta} \biggr\rangle \approx \frac{E_0r_0\gamma_0}{(p_0c/e)\beta_0}\,
\biggl(\biggl\langle \frac{\gamma}{\gamma_0} -1 \biggr\rangle 
+m\,\biggl\langle \frac{x}{r_0} \biggr\rangle 
-\frac{m^2-m}{2}\,\biggl\langle \frac{x^2}{r_0^2} \biggr\rangle \biggr)\,.
\eq
Talman and Talman employ the notation $m$ instead of $n$ for the field index, also $m_p$ for the particle mass.
I have remarked elsewhere \cite{ManeArXivCC14_5_dadt_long}
that the fundamental ideas leading to the above expression are questionable,
but my purpose here is to point out an error of algebra in the above expression,
and to analyze the consequences thereof.

First we need to verify the above formula.
First, eq.~(13) in \cite{TalmanIPAC2012} states
\bq
\label{eq:talmanipac2012_eq13}
\biggl\langle \frac{d\alpha}{d\theta} \biggr\rangle \approx \biggl\langle\frac{eE_0(r_0+x)^2}{(Lc\beta(x)}\biggr\rangle\,
\biggl(\Bigl(\frac{g}{2}-1\Bigr)\langle \gamma\rangle -\frac{g}{2}\Bigl\langle\frac{1}{\gamma}\Bigr\rangle \biggr) \,.
\eq
Next Talman and Talman employ the relativistic virial theorem to deduce (eq.~(16) in \cite{TalmanIPAC2012})
\bq
\biggl\langle\frac{1}{\gamma}\biggr\rangle = \langle\gamma\rangle - \frac{E_0r_0}{m_pc^2/e}\,\biggl\langle\frac{r_0^m}{r^m}\biggr\rangle \,.
\eq
They use this in eq.~\eqref{eq:talmanipac2012_eq13}.
Also the motion is at the magic gamma, so $a=1/(\beta_0^2\gamma_0^2)$, so $g/2 = 1+a = 1/\beta_0^2$.
Then
\bq
\begin{split}
-\biggl\langle \frac{d\alpha}{d\theta} \biggr\rangle &\simeq \biggl\langle\frac{eE_0(r_0+x)^2}{(Lc\beta(x)}\biggr\rangle\,
\biggl( -\Bigl(\frac{g}{2}-1\Bigr)\langle \gamma\rangle 
+\frac{g}{2}\,\langle\gamma\rangle 
-\frac{g}{2}\,\frac{E_0r_0}{m_pc^2/e}\,\biggl\langle\frac{r_0^m}{r^m}\biggr\rangle 
\biggr) 
\\
&\simeq \frac{eE_0r_0^2}{p_0r_0c\beta_0}\,
\biggl( \langle \gamma\rangle 
-\frac{1}{\beta_0^2}\,\frac{m_pc^2\gamma_0\beta_0^2}{m_pc^2}\,\biggl\langle\frac{r_0^m}{r^m}\biggr\rangle 
\biggr) 
\\
&\simeq \frac{E_0r_0\gamma_0}{(p_0c/e)\beta_0}\,
\biggl( \biggl\langle \frac{\gamma}{\gamma_0} \biggr\rangle 
-\biggl\langle\frac{r_0^m}{r^m}\biggr\rangle 
\biggr) \,.
\end{split}
\eq
Next note that
\bq
\frac{r_0^m}{r^m} = 
\frac{1}{(1+x/r_0)^m} = 1 - m\,\frac{x}{r_0} + \frac{m(m+1)}{2}\,\frac{x^2}{r_0^2} + \cdots \,.
\eq
Hence
\bq
\label{eq:talmanipac2012_usevirial}
-\biggl\langle \frac{d\alpha}{d\theta} \biggr\rangle \simeq 
\frac{E_0r_0\gamma_0}{(p_0c/e)\beta_0}\,
\biggl( \biggl\langle \frac{\gamma}{\gamma_0} - 1 \biggr\rangle 
+m\, \biggl\langle \frac{x}{r_0}  \biggr\rangle 
-\frac{\color{red} \bm{m^2+m}}{2}\, \biggl\langle \frac{x^2}{r_0^2} \biggr\rangle 
\biggr) \,.
\eq
Hence there is an error of algebra in the last term in eq.~\eqref{eq:talmanipac2012_eq17}
(i.e.~eq.~(17) in \cite{TalmanIPAC2012});
the coefficient should be $m^2+m$ not $m^2-m$.

\section{Relativistic virial theorem}
Talman and Talman \cite{TalmanIPAC2012} employ the relativistic virial theorem.
The virial is defined as
\bq
G = \bm{r}\cdot\bm{p} \,.
\eq
Talman and Talman state that the time rate of change in bends is (eq.~(15) in \cite{TalmanIPAC2012}) 
\bq
\frac{dG}{dt}\biggr|_{\rm bend} = m_pc^2\gamma - m_pc^2\,\frac{1}{\gamma}
- eE_0r_0\,\frac{r_0^m}{r^m} \,.
\eq 
We can derive this.
Note that $\bm{p}=m_p\gamma \bm{v}$, because there is no vector potential.
Hence
\bq
\begin{split}
\frac{dG}{dt}\biggr|_{\rm bend} &= \frac{d\bm{r}}{dt}\cdot\bm{p} + \bm{r}\cdot\frac{d\bm{p}}{dt}
\\
&= m_p\gamma v^2 - eE_0r_0\,\frac{r_0^m}{r^m} 
\\
&= m_pc^2\gamma - m_pc^2\,\frac{1}{\gamma}
- eE_0r_0\,\frac{r_0^m}{r^m} \,.
\end{split}
\eq 
Talman and Talman state that for bounded motion, one expects the time-averaged rate of change of $G$ to vanish.
(This is standard in the literature.)
Hence Talman and Talman derive (eq.~(16) in \cite{TalmanIPAC2012})
\bq
\label{eq:talmanvirial}
\biggl\langle\frac{1}{\gamma}\biggr\rangle = \langle\gamma\rangle - \frac{E_0r_0}{m_pc^2/e}\,\biggl\langle\frac{r_0^m}{r^m}\biggr\rangle \,.
\eq
Let me employ their notation below.
For the model I treat, we equate the centripetal force to derive (for circular motion)
\bq
\frac{m_pc^2 \gamma \beta^2}{r} = eE_0\,\frac{r_0^{1+m}}{r^{1+m}} \,.
\eq
Hence, using $\beta^2=1-1/\gamma^2$,
\bq
\gamma - \frac{1}{\gamma} = \frac{E_0r_0}{m_pc^2/e}\,\frac{r_0^m}{r^m} \,.
\eq 
This agrees with eq.~\eqref{eq:talmanvirial}.
In fact, for circular orbits $G = \bm{r}\cdot\bm{p}=0$, hence $dG/dt=0$ exactly.
Hence it is reasonable to suppose that, for bounded orbits, the time-averaged rate of change of the virial is zero.

\vfill\pagebreak

\vfill\pagebreak

\vfill\pagebreak

\vfill\pagebreak

\end{document}